\documentclass[prl,notitlepage,twocolumn]{revtex4-1}
\usepackage{amsmath,amsfonts, amssymb, amsthm, dsfont}
\usepackage{yfonts}
\usepackage{bm}
\usepackage{mathrsfs}
\usepackage{graphicx}
\usepackage{verbatim}
\usepackage[colorlinks=true,citecolor=blue]{hyperref}
\usepackage[normalem]{ulem}
\usepackage{empheq}
\usepackage{tensor}
\usepackage{tikz}
\usetikzlibrary{calc}
\usepackage{mathtools}
\usepackage{adjustbox}
\usepackage{natbib}
\usepackage{flushend}
\makeatletter
\def\maketitle{
\@author@finish
\title@column\titleblock@produce
\suppressfloats[t]}
\makeatother
\DeclareMathOperator{\imag}{imag}

\newcommand{\dd}[1]{\operatorname{d}\!#1\,} 
\newcommand{\ee}{\mathrm{e}} 
\newcommand{\ii}{\operatorname{i}} 
\newcommand{\TT}{\mathcal{T}} 
\newcommand{\KK}{\mathcal{K}} 
\newcommand{\HH}{\mathbb{H}} 

\newcommand{\sgn}{\mathrm{sgn}}
\newcommand{\Tr}{\mathrm{Tr}}
\newcommand{\fTr}{\mathrm{fTr}}
\newcommand{\dg}{\dagger}

\newcommand{\ft}{\otimes_{f}}
\newcommand{\wt}[1]{\widetilde{#1}}
\newcommand{\ket}[1]{\lvert#1\rangle}
\newcommand{\bra}[1]{\langle#1\rvert}

\newcommand{\lrangle}[1]{\langle#1\rangle}
\newcommand{\abs}[1]{\left\lvert#1\right\rvert}
\newcommand{\subf}{\mathop{_{\mkern-4mu f}}}

\newcommand{\vket}[1]{\left| #1\right)}
\newcommand{\vbra}[1]{\left( #1\right|}

\newcommand{\IGG}{\mathrm{IGG}}

\renewcommand{\vec}[1]{\boldsymbol{\mathbf{#1}}}

\newcommand{\lag}{\langle}
\newcommand{\rag}{\rangle}

\usetikzlibrary{decorations.pathreplacing,decorations.markings}
\tikzset{middlearrow/.style n args={2}{
        decoration={markings,
            mark= at position #1 with {\arrow{#2}} ,
        },
        postaction={decorate}
    }
}

\hypersetup{colorlinks=true}
\begin{document}

\title{Variational Tensor Wavefunctions for the Interacting Quantum Spin Hall Phase}
\author{Yixin Ma}
\author{Shenghan Jiang}
\email{jiangsh@ucas.ac.cn}
\author{Chao Xu}
\email{xuchao@ucas.ac.cn}
\affiliation{Kavli Institute for Theoretical Sciences, University of Chinese Academy of Sciences, Beijing 100190, China}

\begin{abstract}
   The quantum spin hall (QSH) phase, also known as the 2D topological insulator, is characterized by protected helical edge modes arising from time reversal symmetry.
   While initially proposed for band insulators, this phase can also manifest in strongly-correlated systems where conventional band theory fails.
   To overcome the challenge of simulating this phase in realistic correlated models, we propose a novel framework utilizing fermionic tensor network states.
   Our approach involves constructing a tensor representation of the fixed-point wavefunction based on an exact solvable model, enabling us to derive a set of tensor equations governing the transformation rules of local tensors under symmetry operations.
   These tensor equations lead to the anomalous edge theory, which provides a comprehensive description of the QSH phase.
   By solving these tensor equations, we obtain variational ansatz for the QSH phase, which we subsequently verify through numerical calculations.
   This method serves as an initial step towards employing tensor algorithms to simulate the QSH phase in strongly-correlated systems, opening new avenues for investigating and understanding topological phenomena in complex materials.
\end{abstract}

\maketitle

\emph{Introduction.--}
The discovery of the quantum spin Hall~(QSH) phase\cite{KaneMele2005} has sparked research interest in studying the interplay between symmetry and topology in quantum materials\cite{QiZhang2011,HasanKane2010}.
Initially proposed as a topological band insulator, the QSH phase is characterized by stable properties such as anomalous helical edge modes and topological response to electromagnetic fields\cite{QiHughesZhang2008}.
It has been found that the QSH phase can also be realized as a Mott insulator in strongly-correlated systems, representing an example of interacting fermionic symmetry-protected topological~(SPT) phases\cite{FidkowskiKitaev2010effects,FidkowskiKitaev2011topological,ChenGuWen2011complete,GuWen2014symmetry,GuLevin2014effect,WangGu2018,WangGu2020construction,WangLinGu2017interacting,ChengTantWang2018loop,WangPotterSenthil2014classification,WangSenthil2014interacting,senthil2015symmetry,Witten2016fermion,FreedHopkins2021reflection}.
Solvable models based on commuting-projector Hamiltonians have been used to construct various interacting fermionic SPT phases, including the QSH phase\cite{WareSonChengMishmashAliceaBauer2016,TarantinoFidkowski2016,WangNingChen2018,Metlitski20191d,SonAlicea2019,WangQiFangGu2021exactly}.
However, these models only provide fixed-point wavefunctions and are hardly useful for numerical simulations.

To construct generic variational wavefunctions beyond the fixed point, we turn to fermionic tensor networks\cite{barthel2009contraction,CorbozEvenblyVerstraeteVidal2010,kraus2010fermionic,gu2010grassmann,SchuchGarciaCirac2011,Bultinck2017fermionic,BultinckWilliamsonHaegemanVerstraete2017fermionicmps,wille2017fermionic,cirac2021matrix}.
Our strategy is presented as following.
Motivated by the interacting edge theory, we introduce the fixed-point wavefunction proposed in Ref.~\cite{WangQiFangGu2021exactly}, and then translate it to the tensor network state.
With such tensor representation, we extract a set of tensor equations for symmetry actions on tensors. 
From tensor equations, we obtain algebraic data characterizing the anomalous edge theory of the QSH phase. 
Finally, we apply our method to a spin-1/2 fermionic system on honeycomb and square lattice: by listing and solving tensor equations, we get variational ansatz for the QSH phase on such systems.
We further calculate topological invariants\cite{Shapourian2017many,shiozaki2018many} of the variational wavefunction based on our ansatz to show a parameter region of QSH phase which can be used for numerical simulations.

\emph{Interacting edge theory.--}
The QSH phase hosts charge conservation symmetry generated by $n_f$ and time reversal symmetry $\TT$, where 
\begin{align}
    \TT^2=\exp[\ii\pi n_f]\equiv F~,\quad
    \TT\cdot n_f\cdot\TT^{-1}=n_f~.
    \label{eq:ti_sym_relation}
\end{align}
Here, $F$ is the fermion parity operator.

To get intuition about the interacting bulk wavefunction, we start from its anomalous edge states, which is described by massless helical Dirac fermions: 
\begin{align*}
    H_{edge}&=\int\dd{x} (-\ii v_F)\left[ \psi_R^\dg(x)\partial_x\psi_R(x)-\psi_L^\dg(x)\partial_x\psi_L(x) \right]~,
\end{align*}
where $\psi_{L/R}$ is the left/right moving fermion mode, and $v_F$ the fermion velocity.
$\TT$ acts as $\psi_{R/L}\to\pm\ii\psi_{L/R}$, forbidding mass terms opening a gap.

The interacting edge theory can be analyzed by the bosonization method\cite{haldane1981luttinger}.
Conjugate fields $\phi(x)$ and $\theta(x)$ are introduced, both with periodicity $2\pi$, where $[\partial_x\theta(x),\phi(x')]=2\pi\ii\delta(x-x')$.
With these hydrodynamic variables, $\psi_{R/L}(x)\sim\exp[-(\ii\phi(x)\pm\ii\theta(x)/2)]$, charge density $\delta\rho(x)=-\partial_x\theta(x)/2\pi$, and current density $j(x)=\partial_t\theta(x)/2\pi$. 
Symmetry actions on $\theta$ and $\phi$ are derived from its action on $\psi_{R/L}$, where
\begin{align}
    U(\varphi):{}&\phi\to \phi+\varphi~,\quad \theta\to \theta~;\notag\\
    \TT:{}&\phi\to-\phi~,\quad \theta\to\theta+\pi~,\quad \ii\to-\ii~.
    \label{eq:luttinger_sym_action}
\end{align}
Lagrangian density for the interacting edge theory is\cite{WuBernevigZhang2006,XuMoore2006}
\begin{align}
    \mathcal{L}_{edge}={}&\frac{1}{2\pi}\partial_x\theta\partial_t\phi-\frac{v_F}{4\pi}\left( \frac{1}{K}(\partial_x\theta)^2+ K(\partial_x\phi)^2 \right)\notag\\
    &+\alpha\cos(2\theta-2\theta_0)
    \label{eq:interacting_edge}
\end{align}
where $K$ is the Luttinger parameter, and for the non-interacting case $K=2$.
Due to Eq.~\eqref{eq:luttinger_sym_action}, the most relevant symmetric scattering term is $\alpha\cos(2\theta-2\theta_0)$ with scaling dimension $2K$.
It becomes relevant when $K<1$, driving edge to a gapped phase.
For the classical limit where $\alpha\ll0$, ground states are doubly degenerate, characterized by $\lrangle{\theta}=\theta_0$ and $\lrangle{\theta}=\theta_0+\pi$ respectively.
Note that these two states are exchanged under $\TT$, and thus spontaneously break $\TT$ symmetry. 

Topological defects of such edge symmetry breaking phase host anomalous properties.
We consider a time reversal domain wall at $x_0$, with domains $\lrangle{\theta(x<x_0-\epsilon)}=\theta_0$ and $\lrangle{\theta(x>x_0+\epsilon)}=\theta_0+\pi$, as shown in Fig.~\ref{fig:fermion_decorate_vortex}. 
For region $(x_0-\epsilon,x_0+\epsilon)$, $\theta$ rotate clockwise/counter-clockwise.
Such domain wall carries $\pm1/2$ charge\cite{GoldstoneWilczek1981}, as
\begin{align}
    \int_{x_0-\epsilon}^{x_0+\epsilon}\dd{x}\delta\rho(x)
    =\int_{x_0-\epsilon}^{x_0+\epsilon}\dd{x}\left(-\frac{\partial_x\theta(x)}{2\pi}\right)
    =\pm\frac{1}{2}~.
    \label{}
\end{align}

\emph{The fixed-point wavefunction.-}
We now extend $\theta$-field to bulk. 
The clockwise/counter-clockwise domain wall at edge is identified as $\pm1/2$ vortex, as shown in Fig.~\ref{fig:fermion_decorate_vortex}.
The half-charge edge domain wall motivates a decorated vortex picture\cite{ChenLuVishwanath2014,LiuGuWen2014}: each vortex core carries fermions with $n_f=n_v$, where $n_v$ is the winding number.
Note that $\TT$ symmetry can be recovered by proliferating vortices, and $n_f$ is conserved during this process due to conservation of total vorticity.

\begin{figure}[htpb]
    \centering
    \includegraphics[width=0.45\textwidth]{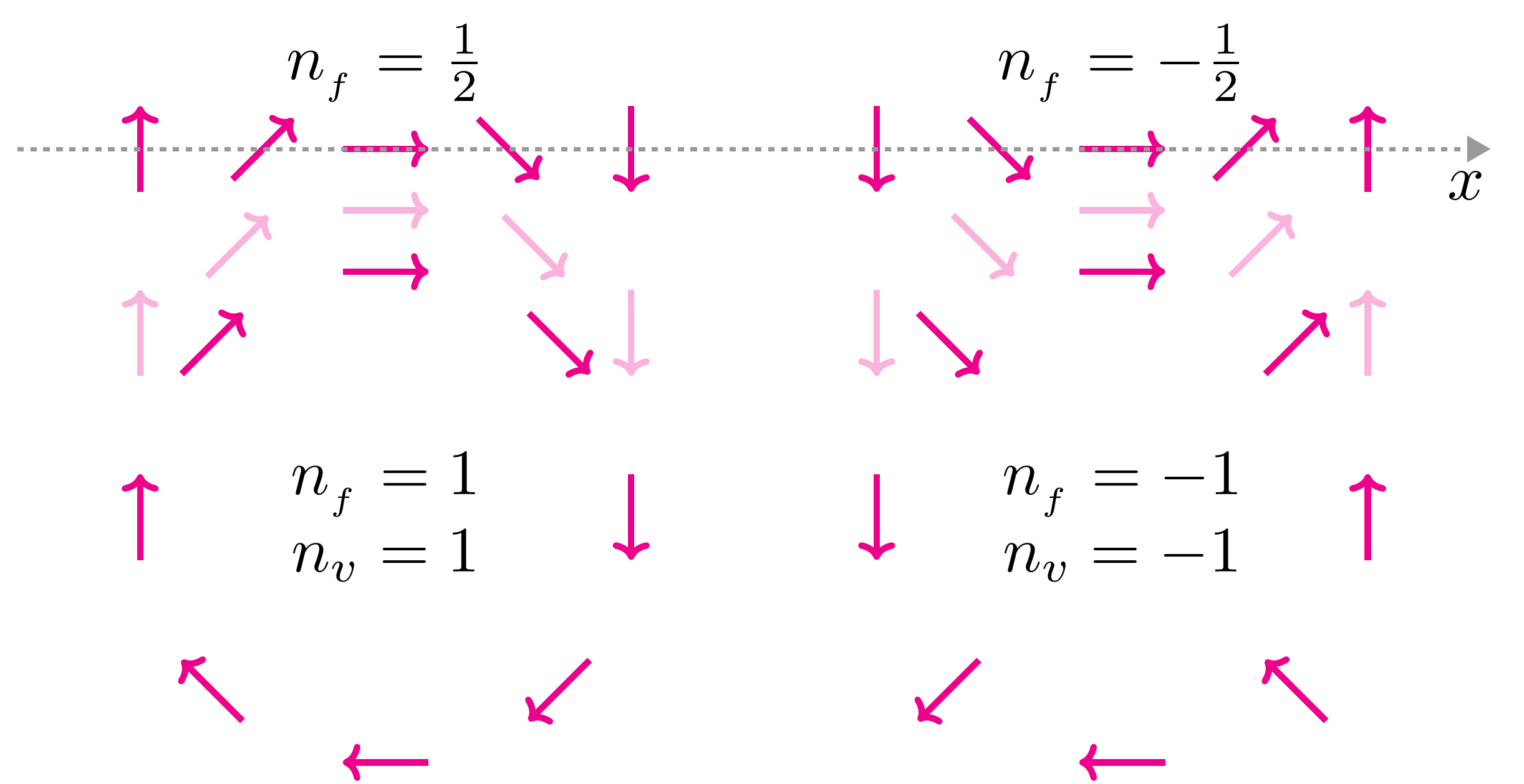}
    \caption{Edge domain walls and bulk vortex of $\theta$-field with fermion decoration.
    }
    \label{fig:fermion_decorate_vortex}
\end{figure}

With the decorated vortex picture, we introduce the fixed-point wavefunction\cite{WangQiFangGu2021exactly}.
As in Fig.~\ref{fig:qsh_fixed_pt_wf}, we consider a system with spin-$1/2$ fermions $f_{\sigma}$'s at a honeycomb lattice, and Ising spins $\ket{\tau}$'s at the dual triangular lattice, where $\sigma,\tau=\uparrow/\downarrow$.
$\TT$ flips both spins:
\begin{align}
    \TT: \ket{\uparrow}\leftrightarrow  \ket{\downarrow}~,~~f_{\sigma}\to \sigma^y_{\sigma\sigma'} f_{\sigma'}~,~~\ii\to-\ii~.
    \label{eq:time_reversal_on_phys_dof}
\end{align}

\begin{figure}[htpb]
    \centering
    \includegraphics[scale=1]{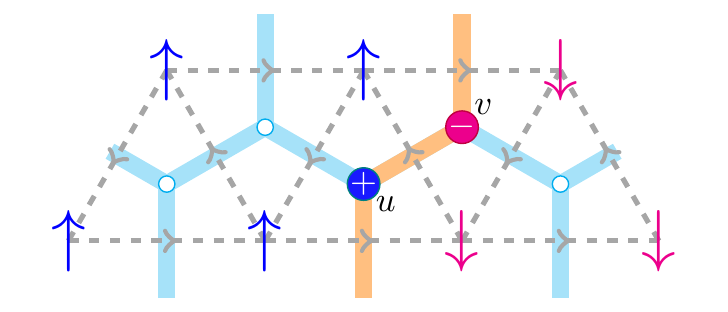}
    \caption{Configuration of the QSH phase's fixed-point wavefunction.
        $f_{\sigma}$ occupy the honeycomb lattice, while $|\tau\rangle$ on the dual lattice.
        Crossing an Ising domain wall along/against oriented bonds of the dual lattice introduces a $\pm\pi$ phase shift.
        Spins of fermions adhere to the majority rule.
    }
    \label{fig:qsh_fixed_pt_wf}
\end{figure}

Here, Ising spins represent $\theta$-field, which rotate $\pm\pi$ when crossing an Ising domain wall along/against the arrow on the bond.
For arrows in Fig.~\ref{fig:qsh_fixed_pt_wf}, an Ising domain wall going through site $(\vec{r},u/v)$ leads to $n_v=\pm1$ at this site.
To match $n_v$, fermions at site $(\vec{r},u/v)$ are holes/electrons:
\begin{align}
    [n_f,f_{\vec{r}s,\sigma}]=-(-1)^{s}f_{\vec{r}s,\sigma}~,
    \label{}
\end{align}
where $(-1)^s=\pm1$ for $s=u/v$. 
The fermion spin is enforced to follow the majority Ising spins at adjacent plaquettes, as shown in Fig.~\ref{fig:qsh_fixed_pt_wf}.
With such majority rule, one can check that for each domain wall loop, number of $f_\uparrow$ differs from number of $f_\downarrow$ by $\pm6$. 
Let $c$ be an Ising spin configuration and $\ket{\psi_c}$ the corresponding decorated fermion state, we have
\begin{align}
    \TT\ket{\psi_c}=(-1)^{N_{dw}(c)}\ket{\psi_{\TT c}}~,
    \label{eq:config_by_t}
\end{align}
where $N_{dw}$ is the number of domain wall loops in $c$.
The fixed-point wavefunction is expressed as\cite{WangQiFangGu2021exactly}
\begin{align}
    \ket{\Psi}=\sum_{c} \Psi(c)\ket{c}\otimes\ket{\psi_c}~,
    \label{eq:fixed_pt_wf_qsh}
\end{align}
where $\Psi(c)=\pm1$ satisfies $\Psi(c)=(-1)^{N_{dw}(c)}\Psi(\TT c)$\footnote{
    We mention that to fully determine $\ket{\Psi}$, we should explicitly write down the fermion order and set the $\pm1$ phase for each $\Psi(c)$
    However, we will not present such information as it is quite complicated and is unnecessary for the following discussion.
}.

\emph{Tensor network representation.-}
Constructing variational wavefunctions beyond Eq.~(\ref{eq:fixed_pt_wf_qsh}) is highly desirable for practical purposes.
In the following, we present a comprehensive framework based on fermionic projected entangled-pair states (fPEPS).
FPEPS are constructed using fermionic tensors, which are quantum states residing in the fermionic tensor product ($\otimes_f$) of physical and internal legs.
The legs with inward/outward arrows correspond to fermionic Hilbert spaces of ket/bra states, respectively.
Fermionic tensor contraction $\fTr$ are implemented by connecting outward and inward internal legs, defined as
\begin{align}
    \fTr[\bra{i}\otimes_f\ket{j}]=(-)^{\abs{i}\abs{j}}\fTr[\ket{j}\otimes_f\bra{i}]=\delta_{ij}
    \label{}
\end{align}
where $(-1)^{\abs{i}}$~($\abs{i}=0/1$) is the fermion parity of $\ket{i}$. 
Physical wavefunctions are obtained by contracting all internal legs.
Site and bond tensors for fPEPS on honeycomb lattice are drawn in Fig.~\ref{fig:honeycomb_triple_line_fpeps}, where all tensors are set to be \emph{parity even} in this paper.
More details about fPEPS are represented in Sec. I of Supplemental Materials (SM)\footnotemark[41].
\footnotetext[41]{See Supplemental Material.}

\begin{figure}[ht]
    \centering
    \includegraphics[width=0.5\textwidth]{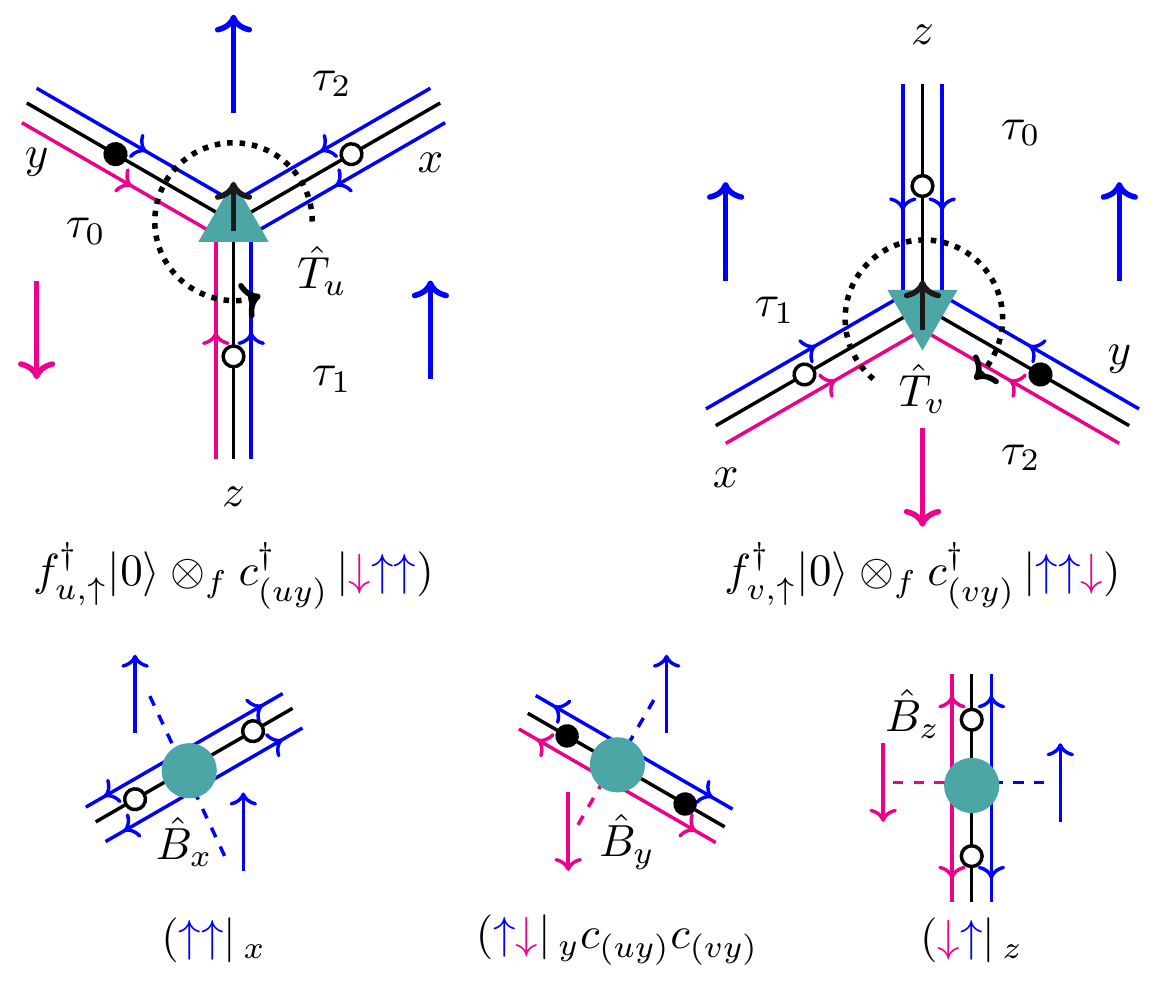}
    \caption{Site tensors $\hat{T}_{u,v}$ and bond tensors $\hat{B}_{x,y,z}$. 
        Physical spin-1/2 fermions live at sites, while physical Ising spins live at bonds.
        Internal legs are represented by triple-lines.
        Lines on two sides are internal Ising spins $\vket{\tau_0\tau_1}$, whose order follows the dashed arcs.
        The middle line is a spinless fermion $c$, where filled/empty circles label filled/empty states.
    }
    \label{fig:honeycomb_triple_line_fpeps}
\end{figure}

Let us work out fPEPS representation of Eq.~(\ref{eq:fixed_pt_wf_qsh}).
Imposing translational symmetry, we focus on tensors in one unit cell, including site tensors $\hat{T}_{u,v}$ and bond tensors $\hat{B}_{x,y,z}$, as in Fig.~\ref{fig:honeycomb_triple_line_fpeps}.
Physical spin-$\frac{1}{2}$ fermions live at sites, while two physical Ising spins live at two sides of bond centers.
Ising spins within a plaquette are enforced to be the same, and thus are effectively plaquette spins.

An internal leg $(s\alpha)$ is represented as a triple-line, pointing from site $s$ to bond $\alpha$, where the middle line is a spinless fermion mode $c_{(s\alpha)}$, while lines at sides are Ising spins.
Basis states are represented as $(c_{(s\alpha)}^\dg)^{n}\vket{\tau_0\tau_1}_{(s\alpha)}$, where vacuum $\vket{0}_{(s\alpha)}$ is omitted for brevity.
$\tau_0\tau_1$ are ordered counter-clockwise/clockwise around the site $u/v$, as indicated by directed dashed arcs in Fig.~\ref{fig:honeycomb_triple_line_fpeps} .

As all spins within a plaquette are the same, an internal spin state of a site tensor reads $\vket{\tau_1\tau_2}_{(sx)}\vket{\tau_2\tau_0}_{(sy)}\vket{\tau_0\tau_1}_{(sz)}$, which is succinctly expressed as $\vket{\tau_0\tau_1\tau_2}$.
Site tensors for Eq.~(\ref{eq:fixed_pt_wf_qsh}) are
\begin{align}
    &\hat{T}_{u}=\ket{0}\otimes_f\Big[ \vket{\uparrow\uparrow\uparrow}+\vket{\downarrow\downarrow\downarrow} \Big] \\
    &+f_{u,\uparrow}^\dg\ket{0}\otimes_f\Big[ c_{(ux)}^\dg\vket{\uparrow\uparrow\downarrow}+c_{(uz)}^\dg\vket{\uparrow\downarrow\uparrow}+c_{(uy)}^\dg\vket{\downarrow\uparrow\uparrow} \Big]\notag\\
    &+f_{u,\downarrow}^\dg\ket{0}\otimes_f\Big[ -c_{(uy)}^\dg\vket{\downarrow\downarrow\uparrow}+c_{(ux)}^\dg\vket{\downarrow\uparrow\downarrow}-c_{(uz)}^\dg\vket{\uparrow\downarrow\downarrow} \Big]\notag\\
    &\hat{T}_{v}=\ket{0}\otimes_f\Big[ \vket{\uparrow\uparrow\uparrow}+\vket{\downarrow\downarrow\downarrow} \Big]\label{eq:site_tensor_ti}\\
    &+f_{v,\uparrow}^\dg\ket{0}\otimes_f\Big[ c_{(vy)}^\dg\vket{\uparrow\uparrow\downarrow}+c_{(vx)}^\dg\vket{\uparrow\downarrow\uparrow}+c_{(vz)}^\dg\vket{\downarrow\uparrow\uparrow} \Big]\notag\\
    &+f_{v,\downarrow}^\dg\ket{0}\otimes_f\Big[ c_{(vx)}^\dg\vket{\downarrow\downarrow\uparrow}-c_{(vz)}^\dg\vket{\downarrow\uparrow\downarrow}+c_{(vy)}^\dg\vket{\uparrow\downarrow\downarrow} \Big]\notag
\end{align}
Similarly, $\bra{\tau_0\tau_1}$ is short for a bond spin state $\bra{\tau_0\tau_1}_{\alpha}\otimes\vbra{\tau_0\tau_1}_{(u\alpha)}\vbra{\tau_1\tau_0}_{(v\alpha)}$.
Bond tensors are expressed as
\begin{align}
    \hat{B}_\alpha=\bra{\uparrow\uparrow}_\alpha + \bra{\downarrow\downarrow}_\alpha + \bra{\downarrow\uparrow}_\alpha - \bra{\uparrow\downarrow}_\alpha c_{(u\alpha)}c_{(v\alpha)}
    \label{eq:bond_tensor_ti}
\end{align}
As we show in Sec. II of SM\footnotemark[41]{}, two $\TT$-related state from contracting fPEPS satisfy Eq.~(\ref{eq:config_by_t}), so it indeed gives the fixed-point wavefunction.

\emph{Tensor equations.-}
Here, we extract symmetry action rules on internal legs for Eq.~\eqref{eq:site_tensor_ti} and \eqref{eq:bond_tensor_ti}, which pave the way for wavefunctions beyond Eq.~(\ref{eq:fixed_pt_wf_qsh}).
We assume that symmetry on physical legs are pushed to gauge transformation on internal legs\cite{Perez2010characterizing}, see also SM\footnotemark[41]{}.
\begin{itemize}
    \item To impose charge conservation, we require all tensors to be charge neutral, which can be realized by assigning $c_{(s\alpha)}$ to carry charge $(-1)^{1-s}$.
        Note that $f_{s,\sigma}$ carries charge $(-1)^s$, and thus
        \begin{align*}
            \left( n_{f;s}+\sum_{\alpha=x,y,z} n_{f;(s\alpha)}\right)\cdot\hat{T}_s
            =\hat{B}_\alpha\cdot \left(\sum_{s=u,v}n_{f;(s\alpha)}\right)=0
            \label{}
        \end{align*}
    \item $\TT$ action on $(s\alpha)$ are set as
        \begin{align}
            &W_{(s\alpha)}(\TT)=\vket{\uparrow\uparrow}_{(s\alpha)}\vbra{\downarrow\downarrow}_{(s\alpha)}+\vket{\downarrow\downarrow}_{(s\alpha)}\vbra{\uparrow\uparrow}_{(s\alpha)} \label{eq:wt} \\
            &+\ii c_{(s\alpha)}^\dg\vket{\uparrow\downarrow}_{(s\alpha)}\vbra{\downarrow\uparrow}_{(s\alpha)}+\vket{\downarrow\uparrow}_{(s\alpha)}\vbra{\uparrow\downarrow}_{(s\alpha)}c_{(s\alpha)}\notag
        \end{align}
        which gives the following symmetric condition:
        \begin{align*}
            \hat{T}_{s}={}&U_s(\TT)\otimes_f W_{(sx)}(\TT)\otimes_f W_{(sy)}(\TT)\otimes_f W_{(sz)}(\TT)\cdot \hat{T}^*_{s}\notag\\
            \hat{B}_{\alpha}={}&V_{(\alpha0)}(\TT)\otimes_f V_{(\alpha1)}(\TT)\cdot\hat{B}^*_{\alpha}\cdot W^{-1}_{(v\alpha)}(\TT)\otimes_f W^{-1}_{(u\alpha)}(\TT)
        \end{align*}
        Here, $U(\TT)$ and $V(\TT)$ are $\TT$-action on physical legs defined in Eq.~(\ref{eq:time_reversal_on_phys_dof}).
        However, as $W(\TT)$'s are not parity even, it may not lead to a symmetric wavefunction.
        In SM\footnotemark[41]{} Sec.~III, we show that this equation contains a hidden Kasteleyn orientation\cite{cimasoni2007dimers,TarantinoFidkowski2016,WareSonChengMishmashAliceaBauer2016,EllisonFidkowski2019}, which gives a $\TT$-symmetric condition, .
    \item Besides the above physical symmetry, such local tensors also host a ``gauge symmetry'':
        \begin{align}
            \Big( n_{\lambda;(s\alpha0)}+n_{\lambda;(s\wt{\alpha}1)} \Big)\cdot \hat{T}_s &=0~,\notag\\
            \hat{B}_\alpha\cdot\Big( n_{\lambda;(u\alpha a)}+n_{\lambda;(v\alpha a)} \Big)&=0~,
            \label{eq:plq_igg}
        \end{align}
        where $n_{\lambda;(s\alpha a)}=(-1)^{s+a}\ket{\downarrow}\bra{\downarrow}$ with $a=0/1$ labelling two side lines of $(s\alpha)$, and $\wt{\alpha}=\alpha-(-1)^s$. 
        $n_\lambda$'s action on all internal legs in a plaquette $p$ imposes the same spin constraint within $p$, and generates a $U(1)$ symmetry.
        We thus get $[U(1)]^{N_p}$ symmetry, where $N_p$ is the number of plaquettes.
        Note that such symmetry acts trivially on physical legs, and is called ``invariant gauge group''~(IGG)\cite{Wen2002,JiangRan2015symmetric,JiangRan2017}, which is related to topological properties of the phase\cite{GuLevinWen2008,schuch2010peps,SchuchGarciaCirac2011}~(see also SM\footnotemark[41]{}).
\end{itemize}

We now extract group relations between $n_f$, $W(\TT)$ and $n_\lambda$, which are coined as \emph{tensor equations} in this work.
Roughly speaking, IGG gives possible action of the identity element on internal legs, and then symmetry on internal legs satisfy Eq.~(\ref{eq:ti_sym_relation}) up to some IGG element\cite{Wen2002,JiangRan2015symmetric,JiangRan2017}.
From Eq.~(\ref{eq:wt}), the commutator between $n_f$ and $\TT$ on internal legs reads
\begin{align}
    W_{(s\alpha)}(\TT)\cdot n_{f;(s\alpha)}\cdot W_{(s\alpha)}^{-1}(\TT)=n_{f;(s\alpha)}+n_{D;(s\alpha)}
    \label{eq:wt_nf_commutator}
\end{align}
where 
\begin{align}
    n_{D;(s\alpha)}&=(-1)^s\left( \ket{\downarrow\uparrow}\bra{\downarrow\uparrow}-\ket{\uparrow\downarrow}\bra{\uparrow\downarrow} \right)\notag\\
    &=n_{\lambda;(s\alpha0)}+n_{\lambda;(s\alpha1)}
    \label{eq:nd_nlambda}
\end{align}
Physically, $n_D$ gives $U(1)$ gauge theory, but due to the decomposition to $n_\lambda$'s, such gauge theory is killed, leading to short-range entangled phase\footnotemark[41]{}.

For group relation $\TT^2=F$, a na\"ive insertion of an IGG element does not give the correct result.
Instead, from Eq.~(\ref{eq:wt}) and (\ref{eq:nd_nlambda}), we have 
\begin{align}
    \exp\left[ \ii\frac{\pi}{2} n_{D;(s\alpha)}^2 \right]\cdot W_{(s\alpha)}(\TT)\cdot W_{(s\alpha)}^*(\TT) =  F_{(s\alpha)}
    \label{eq:wtwt}
\end{align}
In Sec. IV of SM\footnotemark[41]{}, we show that Eq.~(\ref{eq:wtwt}) is indeed consistent with $\TT^2=F$.

The commutator between $W(\TT)$ and $n_{\lambda}$ completes tensor equations:
\begin{align}
    W_{(s\alpha)}(\TT) \cdot n^*_{\lambda;(s\alpha a)} \cdot W_{(s\alpha)}^{-1}(\TT)=-n_{\lambda;(s\alpha a)}+(-)^{s+a}
    \label{eq:t_action_on_nd_nl}
\end{align}

\emph{Edge theories from tensor equations.-}
In the following, we show that edge properties of the QSH phase can be extracted from tensor equations from Eq.~(\ref{eq:wt_nf_commutator}) to Eq.~(\ref{eq:t_action_on_nd_nl}).
The anomalous edge theory is characterized by fusion of two $\TT$-flux\cite{ElseNayak2014classifying}.
To see this, we turn to the edge theory in Eq.~(\ref{eq:interacting_edge}).
By rotating $\theta(x)$ by $2\pi$ angle counter-clockwise within region $[x_0,x_1]$, we get the current density 
\begin{align}
    \int\dd{t}j(x)=\int\dd{t}\frac{\partial_t\theta}{2\pi}=
    \begin{cases}
        0&x\le x_0 \text{ or } x\ge x_1\\
        1& x_0<x<x_1
    \end{cases}
    \label{}
\end{align}
So, a unit charge is pumped from $x_0$ to $x_1$\cite{QiHughesZhang2008fractional,FuKane2006time}. 
According to Eq.~(\ref{eq:luttinger_sym_action}), rotating $\theta$ by $\pi$ on $[x_0,x_1]$ is equivalent to acting $\TT$ on this segment, which creates $\TT$-flux at two ends\cite{ChenVishwanath2014guaging}.
The unit charge pumping due to $2\pi$ rotation of $\theta$ is interpreted as \emph{two $\TT$-flux fuses to an electron/hole}.

We now extract such fusion rule from tensor equations.
As in Fig.~\ref{fig:partial_A}, to obtain edge theory of system $A$, we cut tensors within $A$ from the infinite fPEPS, contract all internal legs within $A$, and obtain a large tensor $\hat{T}_A$.
$\hat{T}_A$ has $L$ boundary legs labeled by index $j\in\partial A=\{1,2,\dots,L\}$, forming Hilbert space $\HH_{\partial A}$.
As shown in Sec. V of SM\footnotemark[41]{}, the edge Hilbert space $\HH_{edge}$ are formed by states in $\HH_{\partial A}$ that are invariant under IGG action.
Let $P_{edge}$ be the projector from $\HH_{\partial A}$ to $\HH_{edge}$. 
In our case, $P_{edge}$ identifies Ising spins belonging to the same plaquette: $\tau_{j+\frac{1}{4}}=\tau_{j+\frac{3}{4}}$, where $\tau_{j\pm\frac{1}{4}}$ are spins at boundary leg $j$.

\begin{figure}[htpb]
    \centering
    \hspace*{-0.1cm}\includegraphics[scale=1.0]{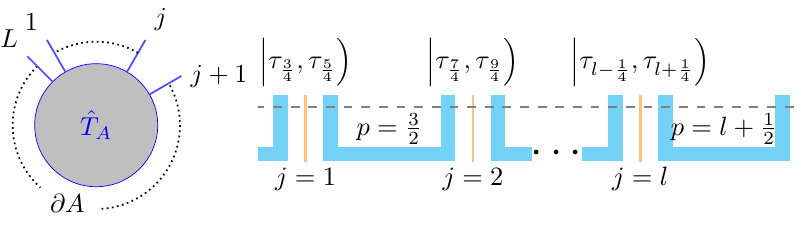}
    \caption{Left: Tensor $\hat{T}_A$ cutting from the infinite fPEPS, whose boundary legs are numbered from $1$ to $L$. 
        Right: Details of boundary legs of $\hat{T}_A$.
        Leg $j$ is a triple-line, representing two Ising spins~(thick blue line) $\tau_{j-\frac{1}{4}}$ and $\tau_{j+\frac{1}{4}}$, and one spinless fermion $c_j$~(thin orange line).
        Plaquette between $j$ and $j+1$ are labeled as $p=j+\frac{1}{2}$. 
    }
    \label{fig:partial_A}
\end{figure}

By projecting $W_{\partial A}(\TT)\equiv\bigotimes_{f;j\in\partial A} W_j(\TT)$ to $\HH_{edge}$, one gets $\TT$ action on edge:
\begin{align}
    U_{edge}(\TT)\mathcal{K}= W_{\partial A}(\TT)\mathcal{K}\cdot P_{edge}
    \label{}
\end{align}
Let $M=\{2,3,\dots,l\}$ be a subregion of $\partial A$. 
$\TT$-flux at ends of $M$ are created by a charge-neutral string operator $U_M(\TT)\KK$, where
\begin{align}
    U_M(\TT)\mathcal{K}=P_{edge}\cdot w_{l+1}\cdot w_{1}\cdot W_M(\TT)\mathcal{K}\cdot P_{edge}
    \label{}
\end{align}
Here, $w_{1/(l+1)}$ are local operators on leg $1/(l+1)$.
The charge-neutral requirement for $U_M(\TT)\KK$ puts the following constraint on $w_{1/(l+1)}$\footnotemark[41]{}
\begin{equation}
    [w_{1},n_{_{f;1}}] = n^{(0)}_{\lambda,\frac{3}{2}}\cdot w_1\,;~
    [w_{l+1},n_{_{f;l+1}}]  = n^{(1)}_{\lambda;l+\frac{1}{2}}\cdot w_{l+1}\,,
    \label{eq:u1f_w_fermion_number}
\end{equation}
where $n_{\lambda;p}^{(0/1)}$ are IGG elements acting on $\vket{\tau_{p\mp \frac{1}{4}}}$. 

Let $j=1$ and $l+1$ be $v$ site, and then we can solve
\begin{align}
    w_1 ={}&\sum_{\tau_{\frac{3}{4}}} c_1 \vket{\tau_{\frac{3}{4}},\downarrow_{\frac{5}{4}}}\vbra{\tau_{\frac{3}{4}},\uparrow_{\frac{5}{4}}} + \vket{\tau_{\frac{3}{4}},\uparrow_{\frac{5}{4}}}\vbra{\tau_{\frac{3}{4}},\downarrow_{\frac{5}{4}}}\,;\notag\\
    w_{l+1} ={}& \sum_{\tau_{l+\frac{5}{4}}} c^{\dagger}_{l+1} \vket{\downarrow_{l+\frac{3}{4}},\tau_{l+\frac{5}{4}}}\vbra{\uparrow_{l+\frac{3}{4}},\tau_{l+\frac{5}{4}}} \notag\\
    &+ \vket{\uparrow_{l+\frac{3}{4}},\tau_{l+\frac{5}{4}}}\vbra{\downarrow_{l+\frac{3}{4}},\tau_{l+\frac{5}{4}}}\,.
    \label{eq:w1wl+1}
\end{align}
It is then straightforward to verify
\begin{align}
    [U_M(\TT)\mathcal{K}]^2 = P_{edge}\cdot \Omega_1\cdot \Omega_{l+1}\cdot \prod_{j=2}^{l}F_j \cdot P_{edge}
\end{align}
where $\Omega_1 =c_1 \exp\left[ \ii \frac{\pi}{2}\cdot n^{(1)}_{\lambda;\frac{3}{2}} \right]$ is a hole, and $\Omega_{l+1} =c_{l+1}^{\dagger} \exp\left[ \ii \frac{\pi}{2}\cdot n^{(0)}_{\lambda;l+\frac{1}{2}} \right]$ an electron (see details in SM Sec. VI\footnotemark[41]{}). 
They can be viewed as quasi-particles from fusing two $\TT$-flux.
We thus recover the anomalous edge theory, which suggests that any fPEPS that satisfies tensor equations belongs to the QSH phase\footnote{
    It is possible that additional IGG elements emerge in the thermodynamic limit\cite{DreyerVanderstraetenChenVerresenSchuch}.
    In such case, fPEPS wavefunction may give other phases, such as spontaneously symmetry breaking phases.
}.

\emph{Variational tensor wavefunctions.-}
The fixed-point wavefunction in Eq.~\eqref{eq:fixed_pt_wf_qsh} is quite artificial, as one needs additional plaquette Ising spins.
In the following, let us try to construct variational wavefunctions for spin-$\frac{1}{2}$ fermions on a honeycomb lattice by solving tensor equations.
We will further demonstrate these wavefunctions give the desired many-body topological invariants in the next part.

We start from fPEPS with two types of site tensors.
Each site tensor has one physical and three internal legs, which can be expressed as $\hat{T}_s=(T_s)_{ijk,p}\vket{i}_{(s1)}\vket{j}_{(s2)}\vket{k}_{(s3)}\ket{p}_s$, with $s=u/v$ labeling the sublattices, and subindices $1$ to $3$ ordering internal legs clockwise. 
Physical spin-$\frac{1}{2}$ fermions $f_{s\sigma}$'s carry opposite charges on site $u$ and $v$.
For simplicity, we choose basis states of an internal leg $(s\alpha)$ to be $\left\{ \vket{\uparrow\uparrow}_{(s\alpha)},~c_{(s\alpha)}^\dg\vket{\uparrow\downarrow}_{(s\alpha)},~\vket{\downarrow\uparrow}_{(s\alpha)},~\vket{\downarrow\downarrow}_{(s\alpha)} \right\}$.

Action of $\TT$ on physical legs follows Eq.~(\ref{eq:time_reversal_on_phys_dof}).
Symmetries impose constraints on site tensor $\hat{T}_s$ as
\begin{align}
    \bigotimes_{\alpha=1}^3 \subf W_{(s\alpha)}(\TT)\otimes_f U_s(\TT) \cdot \hat{T}^{*}_s &= \hat{T}_s\notag\\
    \Big[\sum_{\alpha=1}^3 n_{f;(s\alpha)} + n_{f;s}\Big]\cdot \hat{T}_s &= 0\,.
    \label{eq:teq_example}
\end{align}
In addition, site tensors should also be invariant under plaquette IGG $n_\lambda$'s, as in Eq.~(\ref{eq:plq_igg}).
$W(\TT)$, $n_f$ and $n_\lambda$ satisfy tensor equations from Eq.~(\ref{eq:wt_nf_commutator}) to (\ref{eq:t_action_on_nd_nl}), and can simply take the same form as in the honeycomb example.
By solving these tensor constraints, we obtain $7$ linearly independent solutions for $\hat{T}_{u/v}$, as listed in SM Sec. VIII\footnotemark[41]{}. 
Bond tensors are set to be Eq.~(\ref{eq:bond_tensor_ti}), which satisfy all tensor equations\footnotemark[41]{}.We also calculated the variational ansatz for system in square lattice, see SM\footnotemark[41]{}.

\emph{Many-body topological invariants.-}
One can also diagnose the interacting QSH wavefunction by calculating the many-body topological invariant proposed in Ref.~\onlinecite{shiozaki2018many}.
The system is put on a cylinder periodic in $y-$direction and open in $x-$direction, which is divided as follows
\begin{align*}
    \adjincludegraphics[valign=c,scale=0.3]{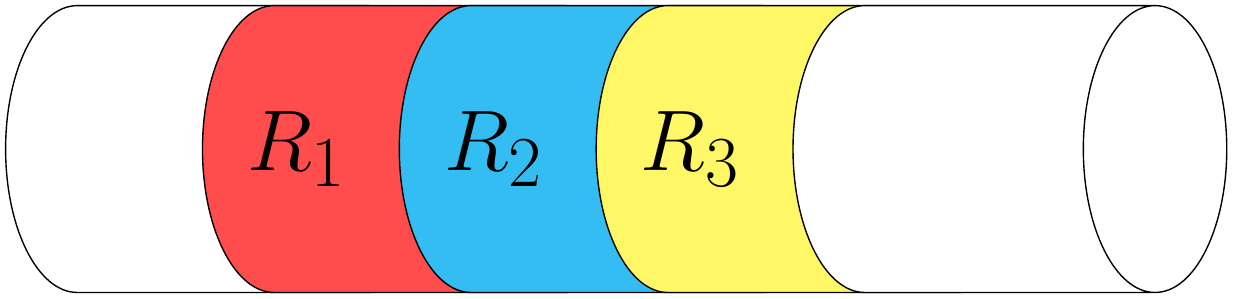}
\end{align*}
The topological invariant is given by 
\begin{align}
    Z&=\Tr\left[ \rho^{+}_{R_{1}\cup R_{3}} C^{R_{1}}_{T}(\rho^{-}_{R_{1}\cup R_{3}})^{\mathsf{T}_{1}}[C^{R_{1}}_{T}]^{\dagger} \right]
\end{align}
with
\begin{align}
    \rho^{\pm}_{R_{1}\cup R_{3}}&=\Tr_{\overline{R_{1}\cup R_{3}}}\left[ \exp{\frac{\pm2\pi \mathrm{i}y\sum_{\mathrm{r}\in R_{2}} n(\mathrm{r})}{L_{y}}} |\Phi\rangle \langle \Phi| \right]\nonumber
\end{align}
Here, $\mathsf{T}_{1}$ is the fermionic partial transpose of region $R_{1}$, and $C^{R_{1}}_{T}$ is defined by $C^{R_{1}}_{T}c_{j\in R_{1}}(C^{R_{1}}_{T})^{\dag}=c^{\dag}_{k\in R_{1}}U(\mathcal{T})_{kj}$. 
The phase of $Z$ equals $0$/$\pi$ when $|\Phi\rangle$ is in trivial/topological phase.
Such quantity can be calculated using numerics, where we present some in Fig.~\ref{fig:sgnZ_cu2}.
\begin{figure}[htpb]
    \centering
    \includegraphics[scale=0.45]{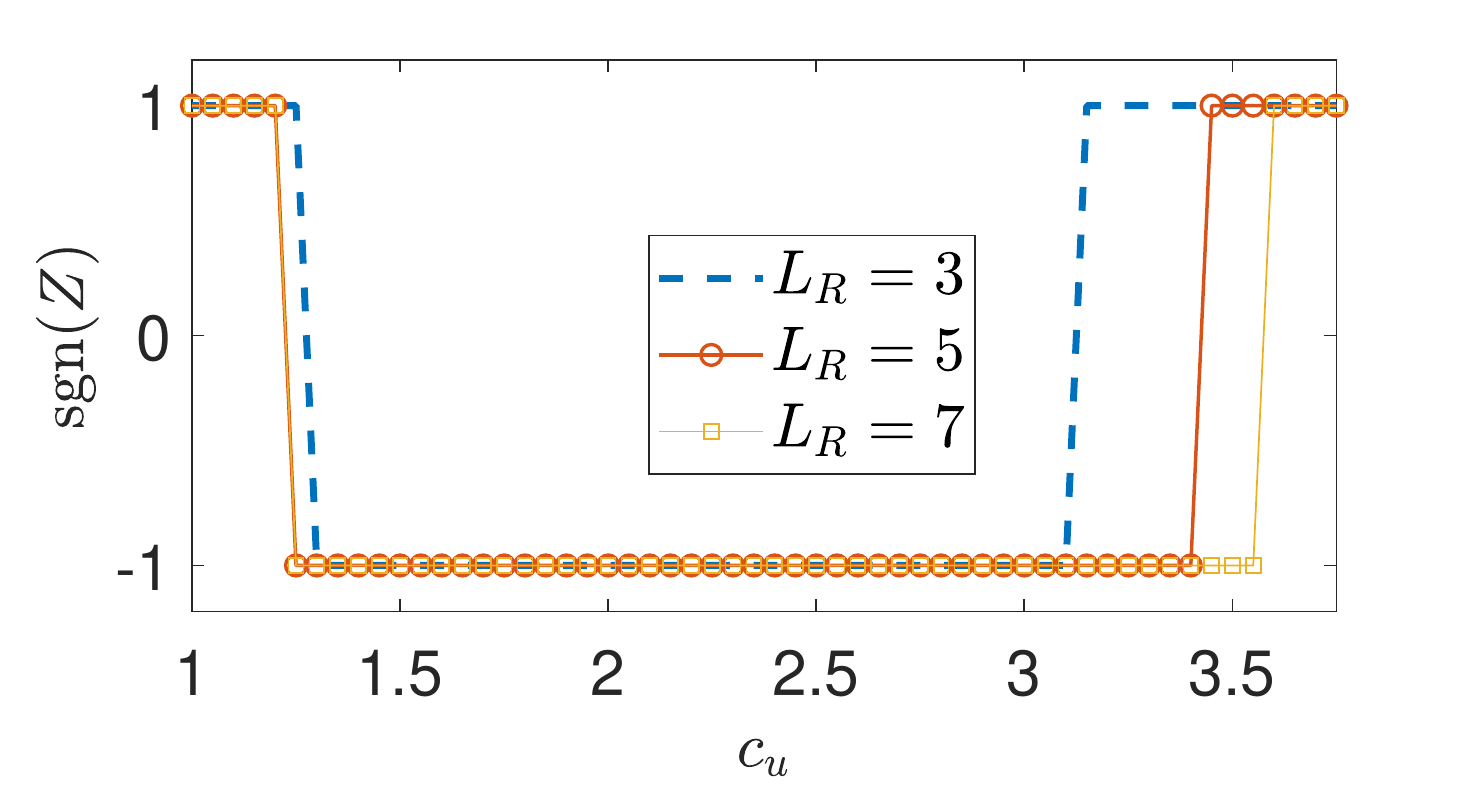}
    \caption{Calculation of $\sgn(Z)$ with respect to $c_u$, where $c_u$ labels some tensor entry, see Supplemental Material for more details.
        $\sgn(Z)=-1$ is a signature of topologically nontrivial phase, see Ref.~\onlinecite{shiozaki2018many}.
        Sizes of $R_{1,2,3}$ are set to be equal, with length in $x$-direction to be $L_R$, and in $y$-direction to be $2$.
    }
    \label{fig:sgnZ_cu2}
\end{figure}
Note that the appearance of $\sgn(Z)=1$ may be due to various reasons, e.g. the finite size effect, see details in SM\footnotemark[41]{}.

\emph{Discussion.-}
In this work, from the fPEPS representation of the fixed-point wavefunction of the QSH phase, we extract tensor equations from Eq.~(\ref{eq:wt_nf_commutator}) to (\ref{eq:t_action_on_nd_nl}) characterizing symmetry rules on local tensors.
By solving these equations, we can obtain general forms for symmetry actions on internal legs. 
Variational ansatz for the QSH phase are solved by imposing such symmetry constraints on local tensors. 

This work leaves several interesting future directions.
Firstly, to express variational ansatz for topological phases in half-filled spin-1/2 \emph{electronic} models, it is necessary to generalize our framework to include tensors with odd parity.
Additionally, developing variational numerical algorithms for symmetric fPEPS wavefunctions obtained in this study would be desirable to simulate the QSH phase in strongly correlated models.
On the analytical side, we aim to explore other fermionic topological phases, such as topological superconductors and topologically ordered phases.
Of particular interest is the investigation of chiral phases, such as the $p+\ii p$ topological superconductor \cite{ReadGreen2000paired,ivanov2001non,StoneRoy2004edge}.
The question of whether fPEPS can accurately represent these chiral phases with a finite bulk gap remains an intriguing puzzle \cite{wahl2013projected,WhalHasslerStefanTuCirac2014symmetries,DubailRead2015tensor}.
Furthermore, tensor networks readily incorporate spatial symmetries \cite{JiangRan2015symmetric}, enabling the construction of variational tensor wavefunctions for gapped electronic liquid phases and high-order topological insulators/superconductors\cite{RasmussenLu2018classification,SongFangQi2018,ElseThorngren2018crystalline}.
Exploring these possibilities holds significant potential for advancing our understanding of exotic topological phases.

We would like to thank Qing-Rui Wang and Xie Chen for helpful discussions.
The work is supported by MOST NO.~2022YFA1403901, NSFC NO.~12104451, and funds from Strategic Priority Research Program of CAS (No.~XDB28000000).

\bibliography{bibfpeps}

\clearpage

\setcounter{secnumdepth}{3}
\setcounter{equation}{0}
\setcounter{figure}{0}
\setcounter{section}{0}
\renewcommand{\thesection}{\Roman{section}}
\renewcommand{\theequation}{S\arabic{equation}}
\renewcommand{\thefigure}{\arabic{figure}}
\newcommand\Scite[1]{[S\citealp{#1}]}
\makeatletter \renewcommand\@biblabel[1]{[S#1]} \makeatother

\title{Supplementary Materials: Variational Tensor Wavefunctions for the Interacting Quantum Spin Hall Phase}
\maketitle
\onecolumngrid

In this supplemental material, we provide a brief review of fermionic tensor network states (Sec.~\ref{app:sym_ftn}), detailed information for the fixed-point tensor network wavefunction of QSH phase (Sec.~\ref{app:fixed_point}), Kasteleyn orientation (Sec.~\ref{app:Kasteleyn}), detailed derivation of $\mathcal{T}^{2}=F$ on internal legs (Sec.~\ref{app:wt2}), edge theory from infinite PEPS (Sec.~\ref{app:edge_peps}), derivation of fusion of $\TT$-flux (Sec.~\ref{app:wtm2}), the variational ansatz for the QSH phase on a square system (Sec.\ref{app:ansatz}), and numerical calculation of many-body topological invariant(Sec.\ref{app:num}).

\section{Symmetric fermionic tensor network states}\label{app:sym_ftn}
In this section, we review the fundamentals of fermionic tensor network states\cite{kraus2010fermionic,BultinckWilliamsonHaegemanVerstraete2017fermionicmps,Bultinck2017fermionic}, and fix our notations used in the main text. 
\subsection{Fermionic tensors and tensor contraction} 
Building blocks of fermionic tensor network states are fermionic tensors, which live in fermionic tensor product~(labeled as $\otimes_f$) of legs.
Legs with inward/outward arrows are local fermion Hilbert spaces for ket/bra states, where the parity of state $\ket{i}/\bra{i}$ is $(-1)^{\abs{i}}$ with $\abs{i}\in\{0,1\}$. 
Exchanging states of two legs gives $-1$ if these two states are both parity odd:
\begin{align}
\ket{i}_a\ft\ket{j}_b=(-1)^{\abs{i}\abs{j}}\ket{j}_b\ft\ket{i}_a
\end{align}

As an example of fermionic tensors, let us consider tensor $\hat{T}$ with three legs, say $a,b,c$:
\begin{align}
    \hat{T}=(T_{abc})_{ijk} \ket{i}_a\ft\ket{j}_b\ft\ket{k}_c
\end{align}
Leg indices $abc$ are sometimes ignored when there is no confusion.
We may also omit $\ft$'s and use a more compact form $\hat{T}=T_{ijk}\ket{ijk}$. 

\begin{figure}[ht]
    \centering
    \includegraphics[scale=1]{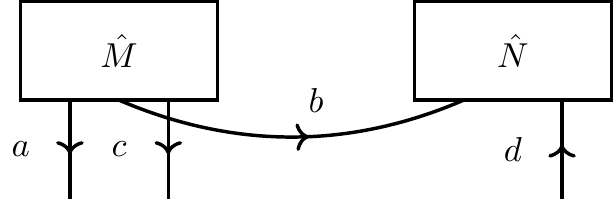}
    \caption{Graphic representation of contraction between fermionic tensors $\hat{M}$ and $\hat{N}$. 
        The outward legs are bra spaces and inward legs are ket spaces.
    The intersection of leg $b$ and $c$ indicates a possible extra minus sign due to exchanging fermions.}
    \label{fig:app_contraction}
\end{figure}

Ket and bra states can be contracted using $\fTr$, defined as
\begin{align}
    \fTr[\bra{i}\otimes_f\ket{j}]=(-)^{\abs{i}\abs{j}}\fTr[\ket{j}\otimes_f\bra{i}]=\delta_{ij}
    \label{eq:ftr}
\end{align}
It is noteworthy that the order of contracted states matters as extra $-1$ may be produced.
Generalization to tensor contractions is straightforward.
As shown in Fig.~\ref{fig:app_contraction}, for two fermionic tensors $\hat{M}=M^{ijk}\bra{i}_a\bra{j}_b\bra{k}_c$, $\hat{N}=N_{lm}\ket{l}_b \ket{m}_{d}$, 
\begin{align}
    \fTr_{b}[\hat{M}\ft \hat{N}]\equiv(-1)^{\abs{j}\abs{k}} M^{ijk}N_{lm}\delta_{jl}\bra{i}\bra{k}\ket{m}~,
\end{align}
which is represented graphically by connecting inward and outward leg.
We may omit $\fTr$'s and $\otimes_{f}$'s and use $\hat{M}\cdot \hat{N}$ to represent tensor contraction. 

For local tensors in tensor network states, there are two types of legs: internal ones and physical ones.
States in physical legs is denoted by $\ket{\bullet}/\bra{\bullet}$, while states in internal legs by $\vket{\bullet}/\vbra{\bullet}$.
To get a physical wavefunction, all internal legs are contracted.
By fixing parity of all local tensors, physical wavefunctions also have fixed parity.
In this work, we focus on the case where all local tensors are \emph{parity even}.

\subsection{Gauge transformation and symmetries of fPEPS}\label{subapp:sym_ftn_gauge_trans}
\begin{figure}
    \centering
    \includegraphics[width=0.8\textwidth]{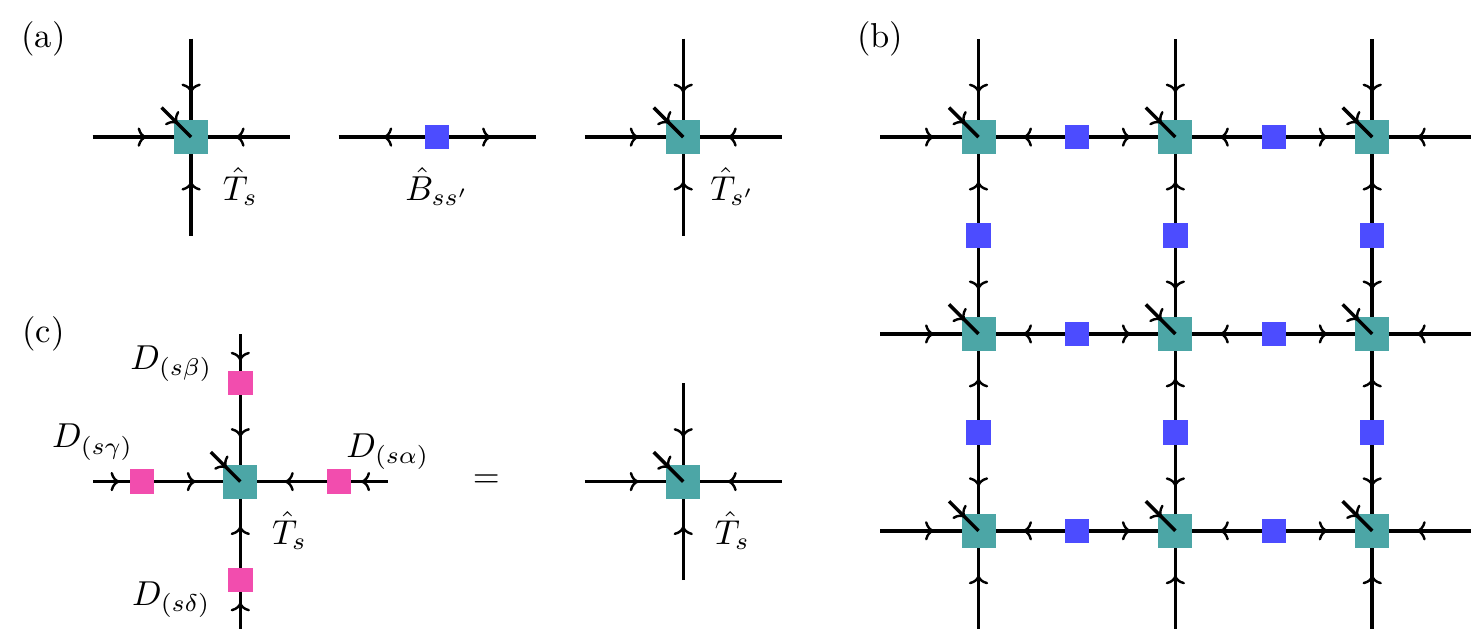}
    \caption{ (a) A bond tensor $\hat{B}_{ss'}$ and its neighbouring site tensors. (b) A $3\times 3$ fPEPS on square lattice with boundary legs. (c) $\IGG$ invariance of a single site tensor. 
    }
    \label{fig:fPEPS}
\end{figure}

We consider a particular type of fermionic tensor network -- fermionic projected entangled-pair state(s)~(fPEPS)\cite{SchuchGarciaCirac2011}.
As shown in Fig.~\ref{fig:fPEPS}(a), we focus on fPEPS with both site tensors $\hat{T}_s$ and bond tensors $\hat{B}_{ss'}$, where $s,s'$ label neighbouring site coordinates.
Without loss of generality, we assume that internal legs of site tensors all point inward~(ket spaces), while those of bond tensors point outward~(bra spaces).
Physical wavefunction then reads $\ket{\Psi}=\fTr\left[ \mathbb{B}\ft \mathbb{T}  \right]$, where $\mathbb{T}=\bigotimes_{s}\subf \hat{T}_s$ and $\mathbb{B}=\bigotimes_{\lrangle{ss'}}\subf\hat{B}_{ss'}$
Note that as all tensors are parity even, different orders of tensor contraction give the same state.

Representation of one physical wavefunction $\ket{\Psi}$ by fPEPS is far from unique.
In particular, different fPEPS represent the same wavefunction if they are related by some gauge transformation, as 
\begin{align}
    \ket{\psi}=\fTr\left[ \mathbb{B}\ft\mathbb{T} \right]
    =\fTr\left[ \mathbb{B}\ft\mathbb{W}^{-1}\ft\mathbb{W}\ft \mathbb{T} \right]
\end{align}
Here, $\mathbb{W}$ and $\mathbb{W}^{-1}$ are tensor products of gauge transformation $W$'s on internal legs:
\begin{align}
    \mathbb{W}&=W_{(s_{1}\alpha_{1})}\ft W_{(s_{1}\alpha_{2})}\ft\cdots\ft  W_{(s_{1}\alpha_{m})} \ft W_{(s_{2}\alpha_{1})}\ft\cdots \ft W_{(s_{n}\alpha_{m})}~,\notag\\
    \mathbb{W}^{-1}&=W^{-1}_{(s_{n}\alpha_{m})}\ft \cdots\ft W^{-1}_{(s_{n}\alpha_{1})}\ft\cdots W^{-1}_{(s_{n-1}\alpha_{m})}\ft\cdots \ft W^{-1}_{(s_{1}\alpha_{1})}
    \label{}
\end{align}
where $(s\alpha)$ labels internal leg, and 
\begin{align}
    W_{(s\alpha)} \vket{i}_{(s\alpha)}=\sum_{b}[W_{(s\alpha)}]_{ji}\vket{j}_{(s\alpha)}~,\quad
    \vbra{i}_{(s\alpha)} W^{-1}_{(s\alpha)}=\sum_{j}\vbra{j}_{(s\alpha)} \left[ W_{(s\alpha)}^{-1} \right]_{ij}
\end{align}
$W$'s in general do not have fixed parity, and thus permuting $W$'s and $W^{-1}$'s may lead to fermion swapping gate.

For the case where $\ket{\Psi}$ is invariant under symmetry $g$, we assume that $g$-action on physical legs can be pushed to gauge transformation on internal legs of local tensors:
\begin{align}
    U_{s}(g)\otimes_f\left( \bigotimes_{\alpha}\subf W_{(s\alpha)}(g) \right)\cdot \hat{T}_{s}=\hat{T}_{s}~,\quad
    \hat{B}_{ss'}\cdot W^{-1}_{(s\alpha)}(g)\ft W^{-1}_{(s'\alpha')}(g)=\hat{B}_{ss'}
\end{align}

The above equations give symmetry constraints for local tensors.
We remark that to get a symmetric wavefunction, orders of $W(g)$'s in the above equation are essential.
In particular, as we will show in Sec.~\ref{app:Kasteleyn}, a valid order of $W(g)$'s gives a Kasteleyn orientation on the lattice. 

As shown in Fig.~\ref{fig:fPEPS}(c), there exists a special kind of gauge transformation $\mathbb{D}$, which leaves every single tensor invariant: 
\begin{align}
    \left(\bigotimes_{\alpha}\subf D_{(s\alpha)}\right) \cdot \hat{T}_{s}=\hat{T}_s~,\quad
    \hat{B}_{ss'}\cdot D^{-1}_{(s\alpha)}\otimes_{f} D^{-1}_{(s'\gamma)} =\hat{B}_{ss'}~.
\end{align}
Such gauge transformation form invariant gauge group~($\IGG$).
In this work, we focus on the case where $D$'s are parity even.

Note that the group always have a trivial center $H$ formed by phase factors $\chi_{(s\alpha)}$ that satisfy $\prod_{(s\alpha)}\chi_{(s\alpha)}=1$. 
In addition, if the $\IGG$ is a $U(1)$ group generated by $n_{D;(s\alpha)}$, namely, $D_{(s\alpha)}(\theta)=\exp[\ii\theta n_{D;(s\alpha)}]$.
We then have $\left(\sum_{\alpha}n_{D;(s\alpha)}\right)\cdot \hat{T}^{s}=0$

\subsection{Plaquette $\IGG$ and vanishing long-range entanglement}\label{subapp:sym_ftn_igg}
In this work, all internal legs can be further decomposed to tensor product of local Hilbert space:
\begin{align}
    \mathcal{H}_{(s\alpha)}=\bigotimes_{a}\subf\mathcal{H}_{(s\alpha a)}
\end{align}
Graphically, an internal leg are represented by multiple lines, and we use index $a$ to label these lines.

\begin{figure}[h]
    \centering
    \includegraphics[width=0.8\textwidth]{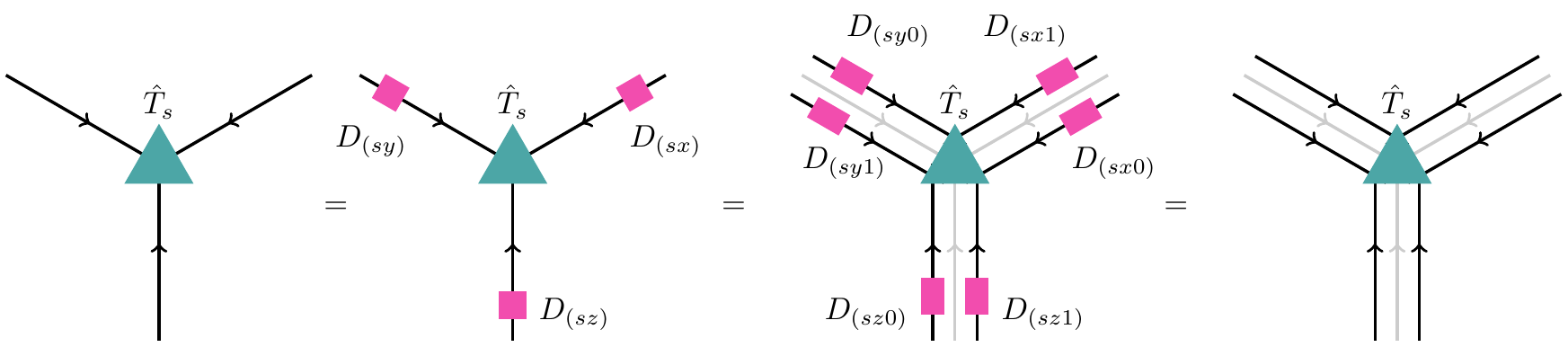}
    \caption{Any $\IGG$ element can be decomposed to plaquette IGG.}
    \label{fig:Plaquette_decomposition}
\end{figure}

As shown in Fig.~\ref{fig:Plaquette_decomposition}, we assume that all elements of $\IGG$ are parity even and have a ``plaquette decomposition''\cite{JiangRan2017}
\begin{align}
    D_{(s\alpha)}=\bigotimes_{a}\subf D_{(s\alpha a)}
\end{align} 
Here, $a=0/1$ denote lines at two sides. 

In particular, for site tensors on honeycomb lattice, the plaquette IGG action gives
\begin{align}
    \adjincludegraphics[scale=0.8,valign=c]{ 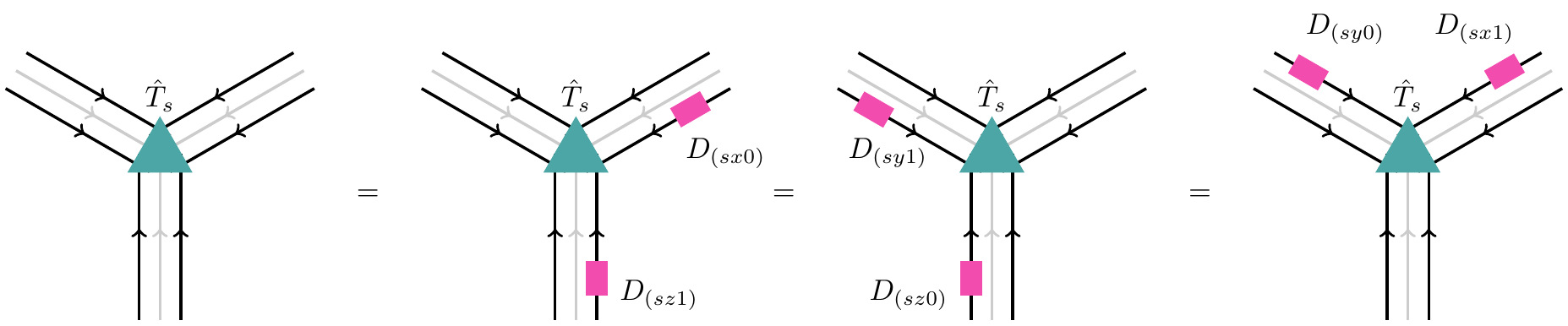}
    \label{eq:plaquette_equation}
\end{align}

Nontrivial $\IGG$ element often leads to topological ground state degeneracy.
To see this, let us consider a fPEPS with periodic boundary condition on square lattice:
\begin{align*}
    \ket{\Psi}=\adjincludegraphics[scale=0.5,valign=c]{ 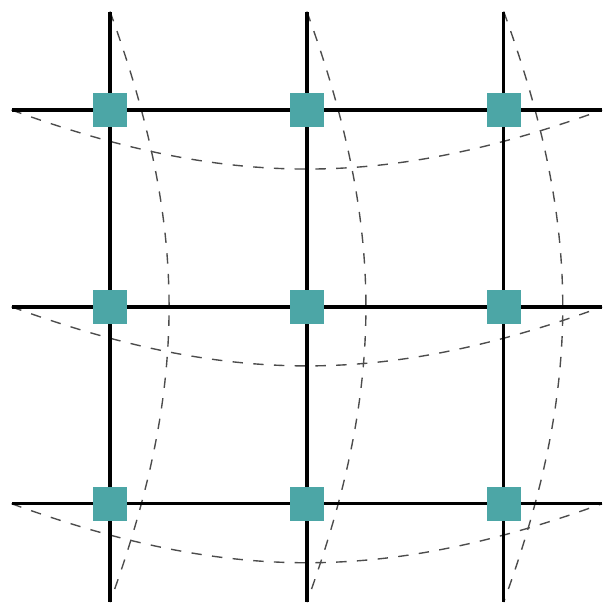}~,
\end{align*}
where the bond tensors, physical legs and leg orientations are omitted for brevity.
If such tensor network has a nontrivial $\IGG$ element, we can insert ``$\IGG$ loops'' in internal legs, which leaves the wavefunction invariant:
\begin{align*}
    \ket{\Psi}=\adjincludegraphics[scale=0.5,valign=c]{ fig/fPEPS_periodic.pdf}~= \adjincludegraphics[scale=0.5,valign=c]{ 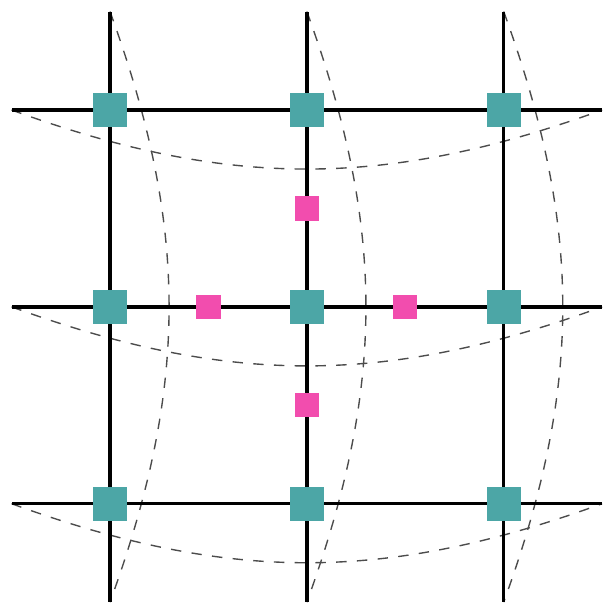}~.
\end{align*}
However, by inserting non-contractible loops of $\IGG$ action, we obtain a different state as
\begin{align}
    \ket{\Psi_{v}}=\adjincludegraphics[scale=0.5,valign=c]{ 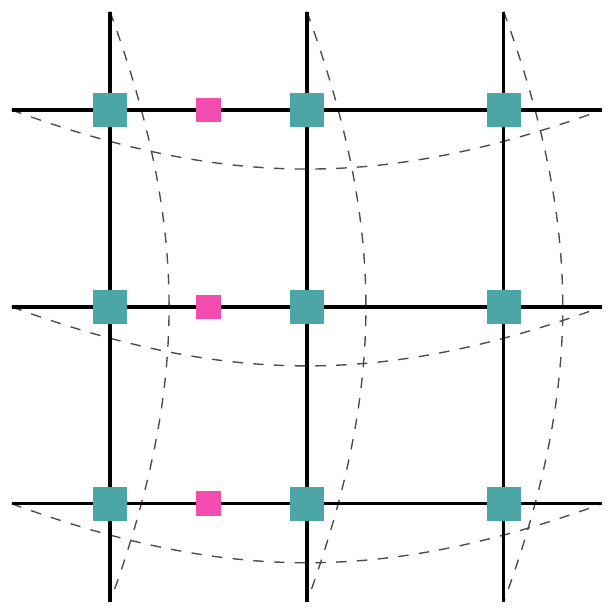}.
    \label{eq:psiv_igg}
\end{align}
As one can move the non-contractible loop without energy cost, $\ket{\Psi_{v}}$ is indistinguishable from $\ket{\Psi}$ by local observables.
Thus, if $\ket{\psi}$ is a ground state of a local Hamiltonian, a nontrivial $\IGG$ indicates that the Hamiltonian has topological ground state degeneracy.
In other words, $\ket{\psi}$ is a long-range entangled state.

However, the plaquette decomposition of the $\IGG$ element kills the long-range entanglement. 
To see this, we apply Eq.~(\ref{eq:plaquette_equation}) for $\ket{\Psi_v}$ in Eq.~(\ref{eq:psiv_igg}), and obtain 
\begin{align*}
    \ket{\Psi_v}=\adjincludegraphics[scale=0.5,valign=c]{ fig/fPEPS_loop_nonc.pdf} ~
    =\adjincludegraphics[scale=0.5,valign=c]{ 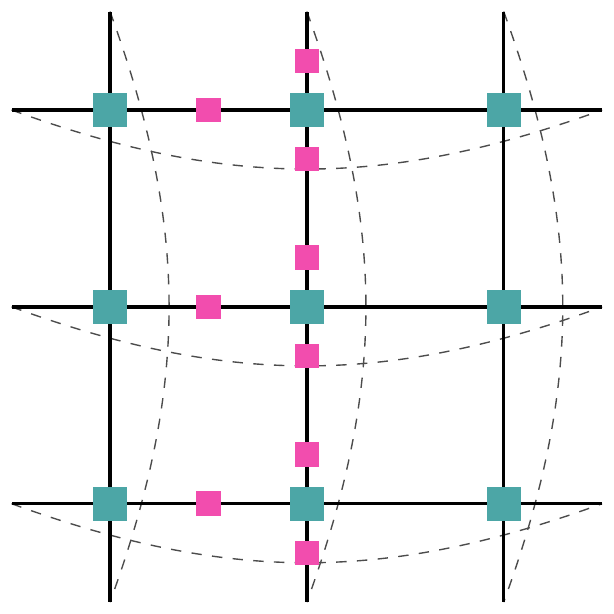}
    =\adjincludegraphics[scale=0.5,valign=c]{ 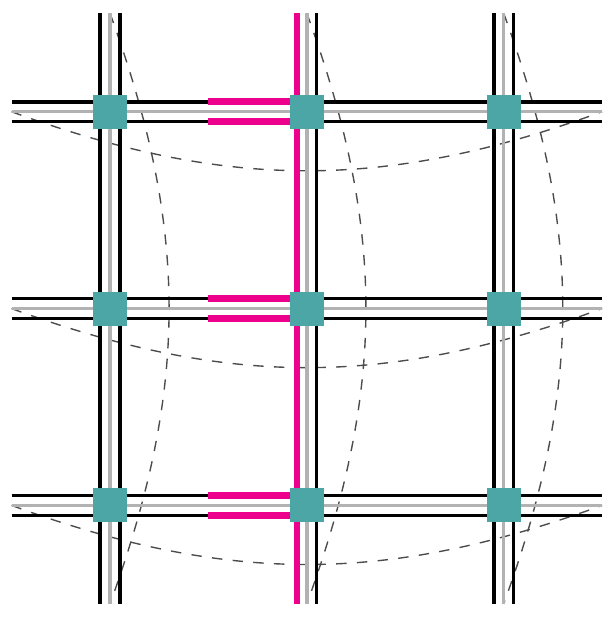}~=\adjincludegraphics[scale=0.5,valign=c]{ 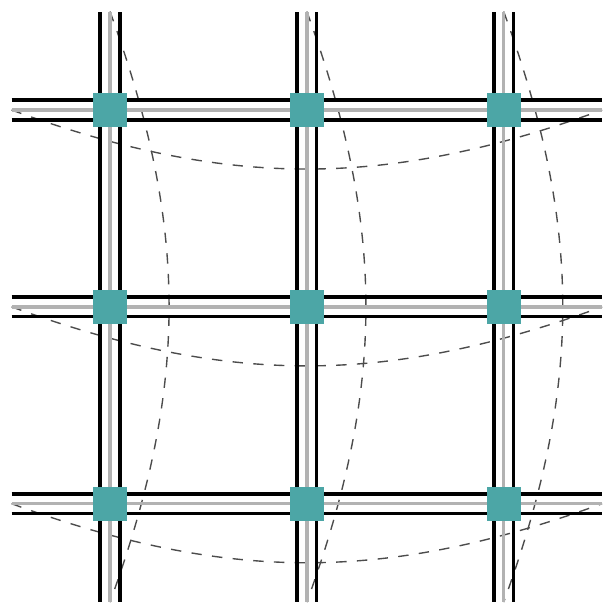}~
    =\ket{\Psi}
\end{align*}
The non-contractible loop of an $\IGG$ element acts trivially on the tensor network state if the $\IGG$ element has plaquette decomposition.
Thus, topological degeneracy is removed and we obtain a short-range entangled state.

\section{Tensor representation of the fixed-point wavefunction for the QSH phase}\label{app:fixed_point}
In this section, we will prove that the wavefunction constructed from tensors in Eq.~(10), (11), and (12) in the main text is $\TT$ symmetric.

The fPEPS physical wavefunction is $\ket{\Psi} = \fTr \left[\mathbb{T}\otimes_f \mathbb{B} \right]$.
Here,
\begin{align}
    \mathbb{T} = \bigotimes_{\vec{r}s}\subf \hat{T}_{\vec{r}s}~,\quad
    \mathbb{B}=\bigotimes_{\vec{r}\alpha}\subf \hat{B}_{\vec{r}\alpha}~,
    \label{}
\end{align}
where $\vec{r}$ is coordinate for unit cell, $s=u/v$, and $\alpha=x/y/z$.
The order of tensors is not important since site and bond tensors are all parity even.

Such wavefunction can be organized according to plaquette Ising spin configurations $c$: 
\begin{align}
    \ket{\Psi}=\sum_c \Psi(c) \ket{c}\otimes\ket{\psi_c}
    \label{}
\end{align}
Here, $c$ represents a Ising spin configuration, and $\ket{\psi_c}$ is the spin-1/2 fermion decoration for $c$. 
We choose some fixed order for physical fermions, and thus there is no ambiguous sign for $\Psi_c$.
Entries of site and bond tensors in the main take $\pm 1$, so $\Psi(c)$ also takes $\pm 1$.
In the following, we will show that for any configuration $c$ and its $\TT$ counterpart $\TT c$, $\Psi(c)=(-1)^{N_{dw}(c)}\Psi(\TT c)$, where $N_{dw}(c)$ is the number of domain wall loops for $c$.
As we argue in Eq.~(7) and (8) in the main text, such state is symmetric under $\TT$.

For the trivial configuration $c$ where all Ising spins points up, $\ket{\psi_c}$ is vacuum state, and $\Psi(c)=\Psi(\TT c)=1$.
Let us consider \emph{configuration $c$ with a single domain wall loop}, and assume this loop contains $2L$ sites and $2L$ bonds.
Overlapping $\ket{\Psi}$ with $\ket{c}$, we obtain $\Psi(c)\ket{\psi(c)}$, which is a new tensor network with site tensors $\hat{T}_{\vec{r}s}^c$ and bond tensors $\hat{B}^c_{\vec{r}\alpha}$.
Any internal leg of this new tensor network only contains one state, which can be either parity even or odd.
Site and bond tensors away from the loop contains no fermion, and is a pure spin state with coefficient $1$.
We focus on tensors in the loop.
We label site tensors along the loop as $\hat{T}_j^c$, where $j\in\{1\cdots 2L\}$ increases clockwise along this loop. 
$\hat{B}^c_{j,j+1}$ labels the bond tensor connecting $j$ and $j+1$.
$\Psi(c)\ket{\psi(c)}$ is then schematically drawn as following:
\begin{align}
    \Psi(c)\ket{\psi(c)}=\adjincludegraphics[scale=1,valign=c]{ 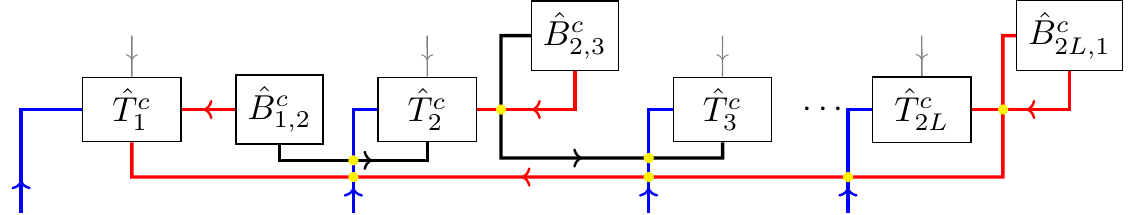}~,
    \label{eq:loop_contraction}
\end{align}
where we only present tensors at the domain wall loop.
Here, blue legs are physical fermions, and light gray legs are internal spins connecting tensors within domains.
Yellow dots are fermion swapping gates. 
Internal legs traveling from right to left are colored red, and give an additional $-1$ when contracting odd parity states according to Eq.~(\ref{eq:ftr}).
$\Psi(\TT c)\ket{\psi(\TT c)}$ can be represented in a similar way.

Coefficients for these configurations come from two contributions: one from swapping fermions, and the other one from the $\pm 1$ entries of tensors.
Let $j=2k-1$ to be $u-$sites, and $j=2k$ to be $v-$sites, we calculate these two contributions respectively in the following.
\begin{itemize}
    \item We assume that the loop encircles an $\uparrow$ domain. 
        From Eq.~(12) in the main text, $\hat{B}_{2k-1,2k}^c$ contains fermion modes $c_{2k-1}$ and $c_{2k}$, while $\hat{B}_{2k,2k+1}^c$ carries zero fermion charge. 
        From Eq.~(\ref{eq:loop_contraction}), contraction between $\hat{B}_{2k-1,2k}^c$ and $\hat{T}_{2k-1}^c$, and contraction between $B_{2k-1,2k}^c$ and $T_{2k}^c$ both give a $-1$.
        Thus, no sign factor is produced.

        Configuration $\TT c$ hosts a $\downarrow$ domain inside the loop.
        $\hat{B}_{2k,2k+1}^{\TT c}$ contains fermion modes $c_{2k}$ and $c_{2k+1}$, while $\hat{B}_{2k-1,2k}^{\TT c}$ is a pure spin state.
        From Eq.~(\ref{eq:loop_contraction}), one concludes that for $k<L$, the contraction between $\hat{B}_{2k,2k+1}^{\TT c}$, $\hat{T}_{2k}^{\TT c}$ and $\hat{T}_{2k+1}^{\TT c}$ contributes $-1$, and contraction between $\hat{B}_{2L,1}^{\TT c}$, $\hat{T}_1^{\TT c}$ and $\hat{T}_{2L}^{\TT c}$ contributes no phase factor.
        So, the sign difference between $c$ and $\TT c$ from fermion contraction is $(-)^{L-1}$.
    \item We now calculate contribution from tensor entries.
        Given site and bond tensors in the main text, one can check that for any loop with length $2L$, under $\TT$ action, there are always $L$ site tensors and $2L$ bond tensors on the loop change signs. 
        This can be seen by noticing that on a domain wall loop, there are always same number of $u-$sites and $v-$sites that locate between domain walls of an $x-$bond and a $y-$bond (or $x-$bond and $z-$bond, or $y-$bond and $z-$bond), while sign factors of these $u-$ and $v-$sites are opposite under $\TT$ action.
        Thus, tensor entries contribute $(-1)^L$ to the sign difference between $c$ and $\TT c$.
\end{itemize}
Combining above contributions, we conclude that $\Psi(c)=-\Psi(\TT c)$ for $c$ with a single domain wall.

For configurations with multiple domain wall loops, one can first moving tensors belonging to a single loop together.
Coefficients for each loop can then be calculated one by one.
So, for any configuration $c$ with $N_{dw}(c)$ domain wall loops, we have $\Psi(\TT c)= (-)^{N_{dw}(c)}\Psi(c)$.

\section{Time reversal symmetry and Kasteleyn orientation}\label{app:Kasteleyn}
When acting $\TT$ on the honeycomb fPEPS, we get $W(\TT)$'s and $W^{-1}(\TT)$'s on internal legs according to Eq.~(14) in the main text.
For any internal leg $(s\alpha)$, $W_{(s\alpha)}(\TT)$ in general are not in the neighbourhood of $W^{-1}_{(s\alpha)}(\TT)$, and one should move them together to cancel each other in the tensor contraction.
Since $W(\TT)$ does not have fixed parity, permuting $W(\TT)$'s and $W^{-1}(\TT)$'s lead to fermion-swap gates.
Eq.~(14) in the main text in general does not lead to a $\TT$-symmetric physical wavefunction due to these swap gates.
In this section, we will show that local $\TT$ constraint gives a symmetric physical wavefunction if and only if a hidden Kasteleyn orientation can be extracted from such local constraint.

Let us first introduce Kasteleyn orientation.
For a given oriented graph, Kasteleyn orientation means that for any face in such graph, the number of clockwise-oriented edges bounding it must be odd.
It has been proven that Kasteleyn orientation exists for any planar graph with an even number of vertices\cite{cimasoni2007dimers}.
The choice of Kasteleyn orientation is far from unique: given an Kasteleyn orientation, one can obtain another one by flipping arrows on all edges connecting a given vertex $v$.
For each vertex, one can choose to flip or not to flip arrows on edges connecting this vertex, and it exhausts all possible choices of Kasteleyn orientation. 
Thus, there are total $2^{N_v}$ choices, where $N_v$ is the total number of vertices.

We now turn to the honeycomb example in the main text. 
Let us explain rules to extract orientation from action of $W(\TT)$'s on local tensors.
We first expand all sites of the honeycomb lattice to triangles as
\begin{align}
    \adjincludegraphics[valign=c]{ 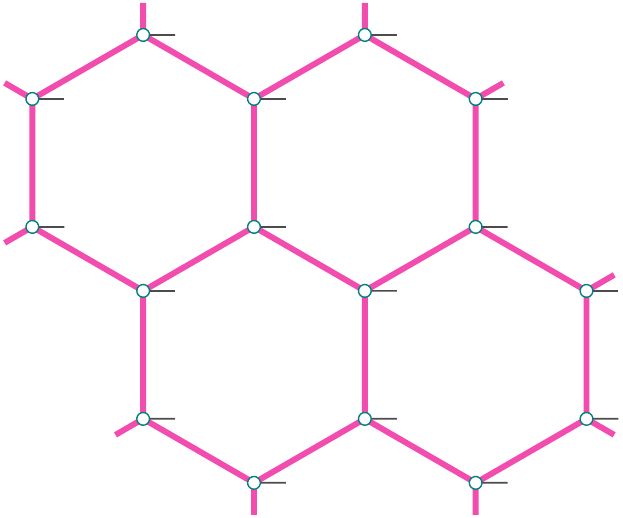}\rightarrow\adjincludegraphics[valign=c]{ 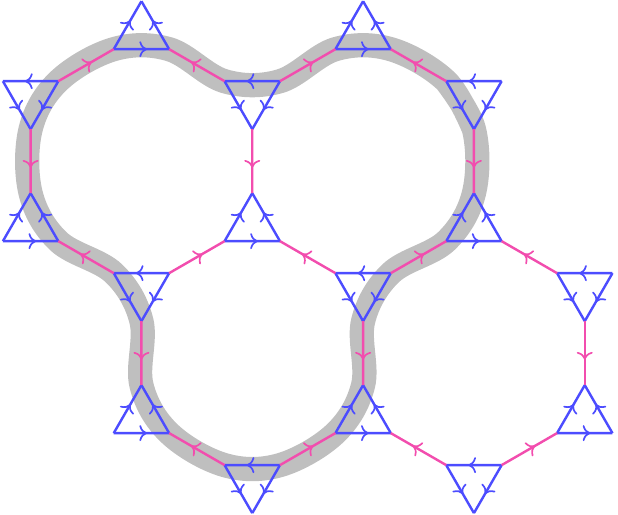}
    \label{eq:honeycomb_to_fisher_kasteleyn}
\end{align}
Vertices in the new lattice live on internal legs of the honeycomb fPEPS, and thus can be labeled as $(s\alpha)$.
Arrows on blue triangles in Eq.~(\ref{eq:honeycomb_to_fisher_kasteleyn}) are extracted from $\TT$-action on site tensors, while arrows on magenta lines in Eq.~(\ref{eq:honeycomb_to_fisher_kasteleyn}) of the original hexagons from $\TT$-action on bond tensors.
More specifically, from $\TT$-symmetric condition in the main text, for $\TT$-action on tensor at site $s$ 
\begin{align}
    U_s(\TT)\cdot\hat{T}_s^* =  \left[ W_{(s\alpha)}(\TT)\otimes_f W_{(s\beta)}(\TT) \otimes_f W_{(s\gamma)}(\TT) \right]^{-1} \cdot \hat{T}_s  \,,
    \label{eq:site_t_action}
\end{align}
the arrows on the corresponding blue triangle is $(s\alpha)\rightarrow(s\beta)$, $(s\beta)\rightarrow(s\gamma)$, and $(s\alpha)\rightarrow(s\gamma)$.
For example,
\begin{align}
    U_s(\TT)\cdot \hat{T}^*_s = \left[ W_{(sx)}(\TT)\otimes_f W_{(sy)}(\TT) \otimes_f W_{(sz)}(\TT) \right]^{-1} \cdot \hat{T}_s  ~~~\Rightarrow \adjincludegraphics[valign=c,scale=2]{ 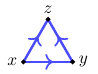}\,.
    \label{eq:site_arrow_rule}
\end{align}
If $\TT$-action on bond tensor $\hat{B}_{\alpha}$ reads
\begin{align}
    V_{(\alpha0)}(\TT)\otimes_f V_{(\alpha1)}(\TT)\cdot \left(\hat{B}_{\alpha}\right)^* 
    =  \hat{B}_{\alpha} \cdot  W_{(s\alpha)}(\TT) \otimes_f  W_{(s'\alpha)}(\TT)\,,
    \label{eq:bond_t_action}
\end{align}
the arrow is drawn from $(s\alpha)$ to $(s'\alpha)$.
Following this rule, $\TT$-symmetry constraint in the main text leads to the oriented graph in Eq.~(\ref{eq:honeycomb_to_fisher_kasteleyn}), and one can check that it indeed gives one Kasteleyn orientation.

In the following, let us prove that the physical wavefunction is $\TT$-symmetric only when the right hand side of Eq.~(\ref{eq:honeycomb_to_fisher_kasteleyn}) is a Kasteleyn orientation.
According to Eq.~(16) in the main text, the commutator between $W(\TT)$ and parity $F$ reads
\begin{align}
    W^*(\TT)\cdot F\cdot \left[W^{-1}(\TT)\right]^*=D\cdot F~,\quad
    \text{where } D=\exp[\ii\pi n_D^*]
    \label{eq:wt_nf_commutator_app}
\end{align}
Here, $D$ can be expressed as $P_e-P_o$, where $P_{e/o}$ are projectors to internal states without/with Ising domain wall.
Using these projectors, we can decompose site and bond tensors to orthogonal sectors, e.g. $P_{e,x}\otimes_f P_{o,y}\otimes_f P_{o,z}\cdot \hat{T}$.
By inserting $P_e+P_o$ on all internal legs of the tensor network wavefunction, the wavefunction equals summation of tensor contractions for  different sectors.
However, as $D$ is an IGG element of the tensor network, local tensors vanish if acted by odd number of $P_o$'s.
So, for the whole tensor network, $P_o$'s must form loops, which gives domain wall loops for physical Ising spins after contraction. 

Given a loop configuration $d$, we act $P_o$'s on internal legs along loops and $P_e$'s on internal legs within domains, and then obtain site tensors $\hat{T}^d$'s and bond tensors $\hat{B}^d$'s.
By contracting internal legs of $\hat{T}^d$'s and $\hat{B}^d$'s, we get physical state $\ket{\psi_d}$. 
Physical wavefunction is obtained by summation over all loop configurations $\ket{\Psi}=\sum_d\ket{\psi_d}$.

In the following, we will prove that $\forall d,~\TT\ket{\psi_d}=\ket{\psi_d}$ if there is hidden Kasteleyn orientation, and thus $\ket{\Psi}$ is invariant under $\TT$.

Let us start from configuration $d$ without any loop.
In this case, we only get $W_e(\TT)$'s when acting $\TT$, which are parity even and free to permute. 
$\ket{\psi_d}$ is apparently invariant under $\TT$: $\TT\ket{\psi_d}=\ket{\psi_d}$.

For configuration $d$ with a single loop, let the number of internal ket legs to be $2L$.
We label internal legs along this loop counter-clockwise by $l$, and assume site tensors sitting between $l=2k-1$ and $l=2k$.
Site tensors along this loop for $\ket{\psi_d}$ can then be named as $\hat{T}_{2k-1,2k}^d$, while bond tensors as $\hat{B}_{2k,2k+1}^d$.
As all tensors are parity even, we rearrange tensors along the loop together in the following way:
\begin{align}
    \ket{\psi_d}=\fTr[\cdots\otimes_f \hat{B}_{2,3}^d\otimes_f \hat{B}_{3,4}^d\otimes_f\cdots\otimes_f \hat{B}_{2L,1}^d\otimes_f \hat{T}_{2L-1,2L}^d\otimes_f \hat{T}^d_{2L-3,2L-2}\otimes_f\cdots \otimes_f \hat{T}_{1,2}^d\otimes_f\cdots]
    \label{eq:psid_tn}
\end{align}
We now act $\TT$ on $\ket{\psi_d}$, and according to Eq.~(\ref{eq:site_t_action}) and Eq.~(\ref{eq:bond_t_action}), it gives $W(\TT)$'s on internal legs of $\hat{T}^d$'s and $\hat{B}^d$'s.
We define $W_{e/o}(\TT)\equiv P_{e/o}\cdot W(\TT)$, which is parity even/odd sector of $W(\TT)$.
From definition of $\ket{\psi_d}$, $W(\TT)$'s act as $W_o(\TT)$'s on internal legs along the loop, while acting as $W_e(\TT)$'s on internal legs away from the loop.
Fermion signs come from permuting $W_o(\TT)$'s and $[W_o(\TT)]^{-1}$'s, and thus we focus on contraction of $W(\TT)$'s along the loop.
We arrange the order of $W_o(\TT)$'s contraction according to Eq.~(\ref{eq:psid_tn}) as
\begin{align}
    \fTr\bigg\{ & \Big( (-1)^{s_{2,3}}\cdot W_{o,2}(\TT)\otimes_f W_{o,3}(\TT) \Big) \otimes_f\cdots\otimes_f \Big( (-1)^{s_{2L,1}}\cdot W_{o,2L}(\TT)\otimes_f W_{o,1}(\TT) \Big)\bigotimes\subf \notag\\
    &\bigg[ \Big( (-1)^{s_{1,2}}\cdot W_{o,1}(\TT)\otimes_f W_{o,2}(\TT) \Big) \otimes_f\cdots\otimes_f \Big( (-1)^{s_{2L-1,2L}}\cdot W_{o,2L-1}(\TT)\otimes_f W_{o,2L}(\TT) \Big) \bigg]^{-1} \bigg\}\notag\\
    ={}&(-1)^{1+\sum_{l}s_{l,l+1}}=1
    \label{eq:psid_wo}
\end{align}
where $s_{l,l+1}=0/1$ if the arrow at $(l,l+1)$ is along/against the direction of the loop (counter-clockwise/clockwise direction).
The last equation is from the definition of Kasteleyn orientation: there are always odd number of arrows against direction of the loop.
So, for configuration $d$ with a single loop, $W(\TT)$'s and $[W(\TT)]^{-1}$'s cancels, and $\ket{\psi_d}$ is $\TT$-symmetric.
In contrast, if the orientation is not Kasteleyn, one can always find a loop configuration $d$, such that the last line of Eq.~(\ref{eq:psid_wo}) gives $-1$, making $\ket{\Psi}$ break $\TT$.

For configuration $d$ with multiple loops, we can arrange all tensors belonging to one loop together, and repeat the above calculation for every loop.
Thus, such $\ket{\psi_d}$ is also $\TT$-symmetric.
In conclusion, $\ket{\Psi}$ is $\TT$-symmetric if and only if the orientation extracted from $W(\TT)$'s is a Kasteleyn orientation.

As we mentioned in the beginning of this part, by flipping arrows on all edges connecting to certain vertices, one gets another Kasteleyn orientation.
In the tensor language, flipping arrows for edges connecting to vertex $(s\alpha)$ corresponds to modifying $W_{(s\alpha)(\TT)}$ to $D_{(s\alpha)}\cdot W_{(s\alpha)}(\TT)$.
To see this, we consider the following $\TT$-action on site tensors:
\begin{align*}
    \left[ D_{(sx)}\cdot W_{(sx)}(\TT)\otimes_f W_{(sy)}(\TT) \otimes_f W_{(sz)}(\TT) \right]^{-1} \cdot \hat{T}_s  
    = \left[  W_{(sy)}(\TT) \otimes_f W_{(sz)}(\TT)\otimes_f W_{(sx)} \right]^{-1} \cdot \hat{T}_s  
    ~~~\Rightarrow \adjincludegraphics[valign=c,scale=2]{ 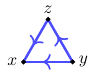}
\end{align*}
Similar logic works for bond tensors.
Following these rules for arrows, there is one-to-one correspondence between gauge transformation $W(\TT)$ and Kasteleyn orientation.

We focus on trivalent lattice in the above argument, where each vertex connects three bonds, and there is at most one domain wall travelling through a vertex.
We now generalize the above argument to generic lattices, where more than one domain walls may meet at sites.

Let us present rules for extracting orientations in generic lattices.
Similar as Eq.~(\ref{eq:honeycomb_to_fisher_kasteleyn}), we first construct a new planar graph, where a site connecting $n$-bonds in the original lattice splits to $n$ vertices in the new graph, and each vertex is labeled by the internal leg index.
Each pair of these $n$ vertices are connected by new edges.
Given $\TT$-action on internal legs, arrows on edges of the new planar graph follows similar rules presented in Eq.~(\ref{eq:site_arrow_rule}) and Eq.~(\ref{eq:bond_t_action}). 
For example, consider site tensor $s$ with four internal legs, arrows can be read from $W(\TT)$ action as
\begin{align*}
    U(\TT) \cdot \hat{T}^*_s  
    = \left[  W_{(sa)}(\TT) \otimes_f W_{(sb)}(\TT)\otimes_f W_{(sc)}\otimes_f W_{(sd)} \right]^{-1} \cdot \hat{T}_s  
    ~~~\Rightarrow \adjincludegraphics[valign=c,scale=2]{ 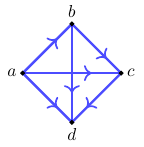}
\end{align*}

To proceed, let us focus on a particular choice of Kasteleyn orientation.
We number the $n$ vertices from $1$ to $n$ clockwise, and let the arrow pointing from $i$ to $j$ if $i<j$.
It is easy to verify that any loop within these $n$ vertices matches the condition for Kasteleyn orientation.
Arrows on bond tensors are chosen to satisfy conditions for Kasteleyn orientation on larger loops.

For configurations without loop crossing, using similar argument presented in the honeycomb case, we conclude that $W(\TT)$'s cancels without additional sign.

\begin{figure}[htpb]
    \centering
    \includegraphics[scale=2]{ 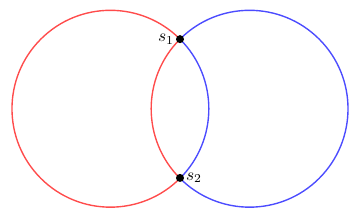}
    \caption{Two domain wall loops~(red and blue) intersect at site $s_1$ and $s_2$.
    Here, $s_{1,2}$ are not ``true crossing points'' of these two loops: they are separate at these two points.}.
    \label{fig:two_loops}
\end{figure}

We consider configuration $d$ where two loops~(colored blue and red) intersect at site $s_1$ and $s_2$.
As shown in Fig.~\ref{fig:two_loops}, colors are chosen such that loops can be separate at these two sites, and there are no ``true crossing points'' between these two loops.

By inserting $P_o$'s on internal legs at these two loops, and $P_e$'s on other internal legs, we obtain tensors $\hat{T}^d$'s and $\hat{B}^d$'s, and wavefunction $\ket{\psi_d}$ from contracting $\hat{T}^d$'s and $\hat{B}^d$'s.
Let the number of internal ket legs of the red/blue loop to be $2L_1/2L_2$.
We label internal legs along these two loops by index $l$, where $1\le l\le 2L_1$ labels internal legs for the red loop and $2L_1+1\le l\le 2L_1+2L_2$ for the blue loop.
Due to the intersecting sites $s_{1/2}$, Eq.~(\ref{eq:psid_tn}) do not directly apply to the case here.
However, at any intersection point, two $W_o^{-1}$'s belonging to one loop are always neighbour to each other~(or can be moved as neighbour without sign).
Thus, we can always move all $W_o^{-1}$'s belonging to one loop together without extra sign, and Eq.~(\ref{eq:psid_wo}) still applies for every single loop.

The above argument can be easily generalized to any loop configurations, as we can always choose loops such that there are no ``true crossing points''.
We conclude that for a $\TT$ symmetric fPEPS of the QSH phase in any lattice, the planar graph extracted from $W(\TT)$'s satisfies Kasteleyn orientation.

\section{$\TT^2=F$ on internal legs}\label{app:wt2}
In this part, we show that Eq.~(18) in the main text is consistent with $\TT^2=F$ when acting on tensors.
We consider a tensor $\hat{T}_A$ with $L$ internal bra legs and physical Hilbert space $\HH_A$. 
According to the main text, the $\TT$-symmetric condition for $\hat{T}_A$ reads
\begin{align}
    U_A(\TT)\cdot \hat{T}_A^*=\hat{T}_A\cdot\bigotimes_{j=1}^{L}\subf W_j(\TT) 
    \label{}
\end{align}
By acting $\TT$ twice, we obtain
\begin{align}
    F_A\cdot \hat{T}_A=U_A(\TT)\cdot U_A^*(\TT)\cdot \hat{T}_A
    =\hat{T}_A\cdot\left(\bigotimes_{j=1}^{L}\subf W_j(\TT) \right)\cdot \left(\bigotimes_{j=1}^{L}\subf W^*_j(\TT) \right)
    \label{}
\end{align}
Due to the unfixed parity of $W(\TT)$'s, one cannot permute $W(\TT)$'s directly.
Instead, from Eq.~(\ref{eq:wt_nf_commutator_app}), we have
\begin{align}
    {}&F_A\cdot \hat{T}_A
    =\hat{T}_A\cdot \left(\bigotimes_{j=1}^{L}\subf W_j(\TT) \right)\cdot \left(\bigotimes_{j=1}^{L}\subf W^*_j(\TT) \right)\nonumber\\ 
    ={}&\hat{T}_A\cdot \bigotimes_{j=1}^{L}\subf (-)^{n_{D,j}\sum_{k<j}n_{D,k} }W_j(\TT)\cdot W_j^{*}(\TT)
    =\hat{T}_A\cdot (-)^{\sum_{k<j}n_{{D,k}}n_{{D,j}} } \bigotimes_{j=1}^{L}\subf W_j(\TT)\cdot W_j^{*}(\TT)\nonumber\\
    ={}&\hat{T}_A\cdot \ee^{-\ii\frac{\pi}{2}\sum_{j\neq k} n_{{D,k}}n_{{D,j}} } \bigotimes_{j=1}^{L}\subf W_j(\TT)\cdot W_j^{*}(\TT)
    =\hat{T}_A\cdot \bigotimes_{j=1}^{L}\subf \ee^{\ii\frac{\pi}{2} n_{{D,j}}^2 } W_j(\TT)\cdot W_j^{*}(\TT)\,,
    \label{eq:wt2_derive}
\end{align}
where the last line comes from the condition that $\sum_{j}n_{{D,j}}=0$ when acting on $\hat{T}_A$.
Then on each internal leg we have 
\begin{align}
    \exp\left[ \ii\frac{\pi}{2} n_D^2 \right]\cdot W(\TT)\cdot W^*(\TT) =  F
\end{align}
or at most differ from $F$ up to an IGG element.
So, by acting $\TT^2=F$ on internal legs of tensors, we get consistent result as Eq.~(18) in the main text. 

\section{Edge theories from infinite PEPS}\label{app:edge_peps}
In this section, we will identify Hilbert space and symmetry action of the edge theory from infinite PEPS.

We cut a finite region $A$ from an infinite PEPS.
By contracting all internal legs within $A$, we obtain a linear map $\hat{T}_A$ from virtual legs at boundary of $A$ -- labeled as $\HH_{\partial A}$ -- to physical legs in the bulk of $A$ -- labeled as $\HH_A$:
\begin{align}
    \hat{T}_A=\sum (T_A)_{i_b i_e}\ket{i_b}\vbra{i_e}~,\quad \ket{i_b}\in \HH_A,~\vbra{i_e}\in \HH_{\partial A}
    \label{}
\end{align}
Here, without loss of generality, we choose all boundary legs to be bra spaces.
For large enough region $A$, $\dim\HH_A\gg \dim\HH_{\partial A}$, so the map can never be surjective.

We can write down a symmetric Hamiltonian for a system on $A$, whose low-energy space is image of $\hat{T}_A$, which is isomorphic to $\HH_{\partial A}/\ker{\hat{T}_A}$. 
As bulk excitations are gapped, low energy states are identified as edge modes, and thus $\HH_{edge}\cong\HH_{\partial A}/\ker{\hat{T}_A}$.

If $\hat{T}_A$ is injective, $\HH_{edge}=\HH_{\partial A}$, and it naturally leads to a tensor product structure of the edge Hilbert space.
If IGG is nontrivial, $\hat{T}_A$ will no longer be injective. 
Given an IGG element whose action on $\partial A$ is $J_{\partial A}$, according to the definition of IGG, $T_A\cdot (\hat{1}_{\partial A}-J_{\partial A})= 0$, and we have $\ker{\hat{T}_A}\supset \imag(\hat{1}_{\partial A}-J_{\partial A})\neq0$.

In this work, we further assume that $\ker{\hat{T}_A}=\left\{ \imag(\hat{1}_{\partial A}-J_{\partial A}) \middle| \forall J\in \IGG \right\}$.
In other words, \emph{IGG determines the edge Hilbert space}:
\begin{align}
    \HH_{edge}=\left\{ \ket{\psi_{\partial A}}~\middle|~J_{\partial A}\ket{\psi_{\partial A}}=\ket{\psi_{\partial A}},~\forall J\in \IGG \right\}
    \label{eq:edge_hibert_space_from_igg}
\end{align}
We define $\hat{T}_A^{-1}:\HH_A\to\HH_{\partial A}$ as pseudo-inverse of $\hat{T}_A$, which satisfies
\begin{align}
    \hat{T}_A\cdot \hat{T}_A^{-1}=P_l~,\quad
    \hat{T}_A^{-1}\cdot \hat{T}_A=P_{edge}~,
    \label{eq:ta_inv_def}
\end{align}
where $P_l$ is the projector from $\HH_A$ to $\imag\hat{T}_A$, while $P_{edge}$ is the projector from $\HH_{\partial A}$ to $\HH_{edge}$.

In the following, let us work out how symmetries act on edge.
Here, we focus on onsite symmetry group $G$.
For $g\in G$, we have 
\begin{align}
    U_A(g)\KK^{s(g)}\cdot T_A=T_A\cdot W_{\partial A}(g) \KK^{s(g)}
    \label{}
\end{align}
So, it is natural to identify $U_{edge}(g)\KK^{s(g)}\equiv P_{edge}\cdot W_{\partial A}(g)\KK^{s(g)}\cdot P_{edge}$ as symmetry action on $\HH_{edge}$.

Note that $\forall J\in \IGG$, we have
\begin{align}
    U_A(g)\KK^{s(g)}\cdot\hat{T}_A
    =\hat{T}_A\cdot W_{\partial A}\KK^{s(g)}
    =U_A(g)\KK^{s(g)}\cdot\hat{T}_A\cdot J_{\partial A}
    =\hat{T}_A\cdot W_{\partial A}\KK^{s(g)}\cdot J_{\partial A}
    \label{}
\end{align}
Together with Eq.~(\ref{eq:edge_hibert_space_from_igg}), we conclude
\begin{align}
    U_{edge}(g)\KK^{s(g)}
    =P_{edge}\cdot W_{\partial A}(g)\KK^{s(g)}
    =W_{\partial A}(g)\KK^{s(g)}\cdot P_{edge} 
    \label{eq:edge_sym_proj_commute}
\end{align}

\section{Fusion of $\TT-$flux}\label{app:wtm2}
As in Sec.~\ref{app:edge_peps}, let us consider a region $A$ described by a large tensor $\hat{T}_A$ with physical Hilbert space $\HH_A$ and boundary legs $\HH_{\partial A}$, where legs at $\partial A$ are labeled by $j=\{1,2,\dots,L\}$.

We create $\TT$-flux at $j=1$ and $l+1$ by inserting a charge-neutral operator $U_M(\TT)\KK\equiv P_{edge}\cdot w_{l+1}\cdot w_{1}\cdot W_M(\TT)\KK\cdot P_{edge}$, where $M=\{2,3,\dots,l\}\in\partial A$. 
Here, $w_{1/(l+1)}$ are local operators at the ends and 
$$W_M(\TT) = \bigotimes_{j=2}^{l}\subf W_j(\TT)$$
We choose $w_{1/(l+1)}$ such that $w_{l+1}\cdot w_{1}\cdot W_M(\TT)\KK$ commute with $P_{edge}$.

To make $U_M(\TT)K$ charge neutral, we require commutator between $w_{1/(l+1)}$ and $n_f$ to be
\begin{align}
     [w_1,n_{_{f;1}}] = {n_{\lambda;\frac{3}{2}}^{(0)}}\cdot w_1~;\quad 
     [w_{l+1},n_{_{f;l+1}}] = {n_{\lambda;l+\frac{1}{2}}^{(1)}}\cdot w_{l+1}
\end{align}
The charge neutral condition is satisfied as $[W_j(\TT),n_{_{f;j}}] = n_{D;j}\cdot W_j(\TT)$ and $\left( n_{\lambda;\frac{3}{2}}^{(0)} + n_{\lambda;l+\frac{1}{2}}^{(1)} + \sum_{j=2}^{l} n_{D,j} \right)\cdot P_{edge} = 0$.

We now calculate fusion of two $\TT$-flux.
Without loss of generality, we assume that $j=1$ and $j=l+1$ legs belong to $v$-sublattice.
By performing similar derivation presented in Eq.~(\ref{eq:wt2_derive}), we have
\begin{align}
    &U_M(\TT)\cdot U_M^{*}(\TT)\\ \nonumber
    = &P_{edge}\cdot \ee^{\ii \pi n_{\lambda;\frac{3}{2}}^{(0)}} \ee^{\ii\frac{\pi}{2} \left(n_{\lambda;l+\frac{1}{2}}^{(1)}\right)^2}w_{l+1}\cdot w_{l+1}^{*}\otimes_f \ee^{\ii\frac{\pi}{2} \left(n_{\lambda;\frac{3}{2}}^{(0)}\right)^2}w_1\cdot w_1^{*}\bigotimes_{j=2}^{l}\subf \ee^{\ii\frac{\pi}{2} n_{D,j}^2 } W_j(\TT)\cdot W_j^{*}(\TT)\cdot P_{edge}\,.
\end{align}
As $\left(n_{\lambda;l+\frac{1}{2}}^{(1)}\right)^2=n_{\lambda;l+\frac{1}{2}}^{(1)}$ and $ \left(n_{\lambda;\frac{3}{2}}^{(0)}\right)^2=-n_{\lambda;\frac{3}{2}}^{(0)}$ for $v-$sites, we obtain
\begin{align}
    (U_M(\TT)\KK)^2 = P_{edge}\cdot \Omega_{l+1} \Omega_1\cdot \prod_{j=2}^{l} F_j \cdot P_{edge}\,,
\end{align}
Here, 
\begin{align}
    \Omega_{l+1} =&{} P_{edge}\cdot \ee^{\ii\frac{\pi}{2} n_{\lambda;l+\frac{1}{2}}^{(1)}}w_{l+1}\cdot w_{l+1}^{*}  \cdot P_{edge} = \ee^{\ii\frac{\pi}{2} n_{\lambda;l+\frac{1}{2}}^{(1)}} c^{\dagger}_{l+1}\,,\\
    \Omega_1 =&{} P_{edge}\cdot \ee^{\ii\frac{\pi}{2} n_{\lambda;\frac{3}{2}}^{(0)}}w_{1}\cdot w_{1}^{*}  \cdot P_{edge} = \ee^{\ii\frac{\pi}{2} n_{\lambda;\frac{3}{2}}^{(0)}} c_1\,.
\end{align}
where we use Eq.~(24) in the main text to fix the final form of $\Omega_{1/(l+1)}$.

\section{Variational ansatz for the QSH phase on square lattice spin-1/2 electronic system}\label{app:ansatz}
In this section, we present detailed derivation for solving tensor equations on a spin-$\frac{1}{2}$ fermionic system on the bipartite square lattice.
Each site tensor has four internal legs and one physical spin-$\frac{1}{2}$ fermion, which is represented as
\begin{align*}
    \adjincludegraphics[valign=c]{ 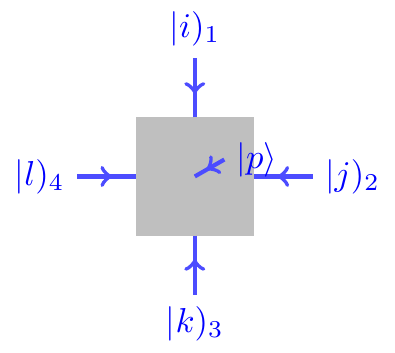}=T_{ijkl,p}\vket{i}_1\vket{j}_2\vket{k}_3\vket{l}_4\ket{p}\,.
\end{align*}
Sites on the bipartite square lattice can be grouped to two types, where we use $\hat{T}_{u/v}$ to label site tensors on different sublattices.
Physical legs are spin-1/2 fermions $f_\sigma$, and charges carried by $f_\sigma$'s are opposite in $u-$ and $v-$site. 
As in the honeycomb example, an internal leg $(s\alpha)$ can also be represented by a triple-line, where the middle line is a spinless fermion mode $c_{(s\alpha)}$, and two side lines are Ising spins $\vket{\tau}_{(s\alpha a)}$, where $s=u/v$, $\alpha\in\{1,2,3,4\}$ and $a=0/1$. 
Orders of $a$ is counter-clockwise/clockwise for site $u/v$.
For simplicity, we focus on four of eight basis states: $\{ \vket{\uparrow\uparrow}_{(s\alpha)},~c_{(s\alpha)}^\dg\vket{\uparrow\downarrow}_{(s\alpha)},~\vket{\downarrow\uparrow}_{(s\alpha)},~\vket{\downarrow\downarrow}_{(s\alpha)}\}$.

Symmetry constraints on site tensor $\hat{T}_s$ are
\begin{align}
    \bigotimes_{\alpha=1}^4\subf W_{(s\alpha)}(\TT)\otimes_f U_s(\TT) \cdot \hat{T}^{*}_s &= \hat{T}_s\notag\\
    \Big[\sum_{\alpha=1}^4 n_{f;(s\alpha)} + n_{f;s}\Big]\cdot \hat{T}_s &= 0
   \label{eq:teq_example2}
\end{align}
where $W(\TT)$, $n_f$, and $n_\lambda$ take the same form as those in the honeycomb example.
To take care of sign factors when acting $W(\TT)$'s on site tensor $\hat{T}_s$, a Jordan-Wigner string $J = \exp \{\ii\pi n_f\}$ is introduced. 
Then,  the ``bosonized'' $W(\TT)$ is
\begin{align}
   W^b_{(sj)}(\TT) = P_e\cdot W_{(sj)}(\TT)+\bigotimes_{k=1}^{j-1}J_k~\otimes P_o\cdot W_{(sj)}(\TT)\,,
\end{align}
$\TT$ symmetry constraint on a site tensor gives
\begin{equation}
   \prod_{\alpha=1}^4 W_{(s\alpha)}^b(\TT)\cdot U(\TT) \cdot T^{*}_s = T_s
   \label{eq:teq_example_boson}
\end{equation}
Here $T_s$ without hat is a ``bosonic tensor'' with entries $(T_s)_{\alpha\beta\gamma\delta,p}$.

Site tensor $\hat{T}_s$ should also satisfy the plaquette IGG condition:
\begin{align}
    \Big( n_{\lambda;(s\alpha0)}+n_{\lambda;(s\wt{\alpha}1)} \Big)\cdot \hat{T}_s = 0,~\forall \alpha\,.\label{eq:iga_example2}
\end{align}
where $\wt{\alpha}=\alpha+(-1)^s$.
This equation identifies internal states within a plaquette:
\begin{align*}
    \adjincludegraphics[valign=c]{ 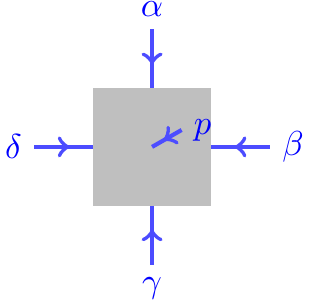}\to\adjincludegraphics[valign=c]{ 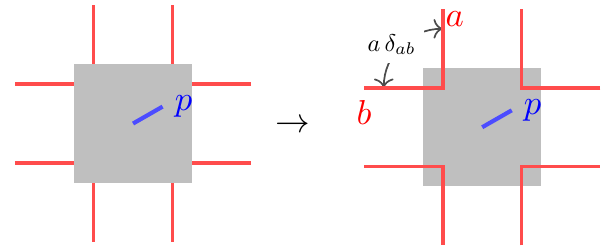}\,.
\end{align*}
The dimension of the tensor $T_s$ is 1024, and after solving the tensor equations in Eq.~\eqref{eq:teq_example2} and \eqref{eq:iga_example2} there are only 14 linearly independent solutions. 
And the solution for $\hat{T}_v$ reads
\begin{align}
    \hat{T}_v = \sum_l^{14} c_l \hat{t}_l\,,
\end{align}
where $c_l$'s are real numbers, and  $\hat{t}_l$ can be represented graphically as
\begin{align*}
    &\adjincludegraphics[valign=c]{ 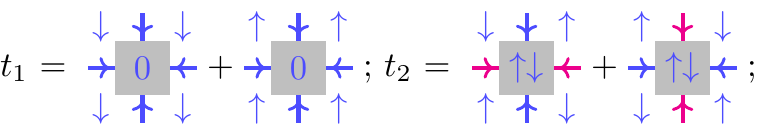}~\adjincludegraphics[valign=c]{ 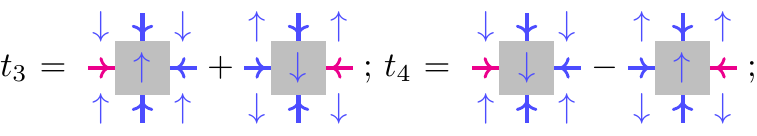}\\
    &\adjincludegraphics[valign=c]{ 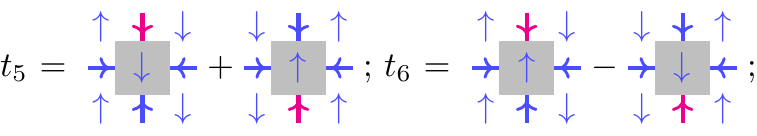}~\adjincludegraphics[valign=c]{ 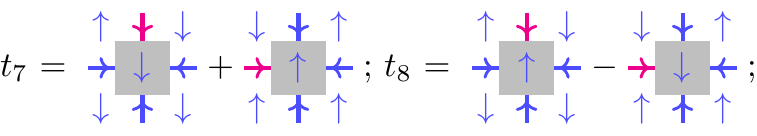}\\
    &\adjincludegraphics[valign=c]{ 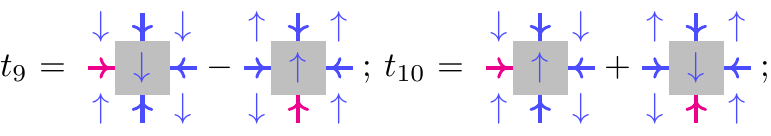}~\adjincludegraphics[valign=c]{ 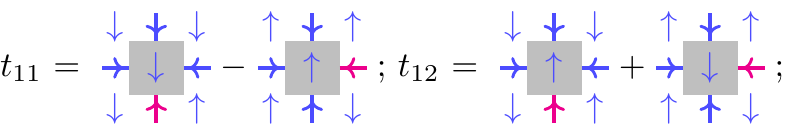}\\
    &\adjincludegraphics[valign=c]{ 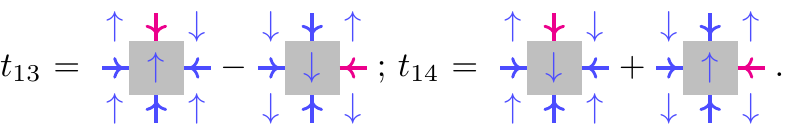}
\end{align*}
Magenta arrow indicates a fermion on the internal leg, and the $\uparrow/\downarrow$ at 4 corners are the identified internal states within the same plaquette.
$\hat{T}_u$'s solution is given by flipping all plaquette Ising spins of $\hat{T}_v$.

Now, let us discuss the bond tensors.
As shown in Sec.~\ref{app:Kasteleyn}, to make the tensor network $\TT$ symmetric, $\TT$ action on bonds should be chosen to satisfy Kasteleyn orientation:
\begin{align}
    \hat{B}_{\alpha}\cdot W_{(u\alpha)}(\TT)\otimes_f W_{(v\beta)}(\TT)&= \hat{B}_\alpha^*
\end{align}
Here, we impose rotational symmetry, and thus all four types of bond tensors share the same form:
\begin{align}
    \adjincludegraphics[valign=c]{ 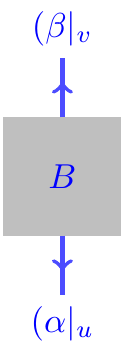}\,=\,\adjincludegraphics[valign=c]{ 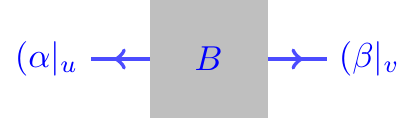}\,=\,\adjincludegraphics[valign=c]{ 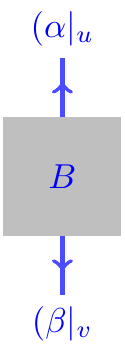}\,=\,\adjincludegraphics[valign=c]{ 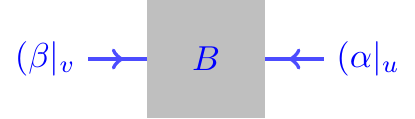} = B_{\alpha\beta} \vbra{\alpha}\vbra{\beta}
\end{align}

Bond tensors should also be invariant under plaquette IGG:
\begin{align}
    \Big( n_{\lambda;(u\alpha a)}+n_{\lambda;(v\alpha a)} \Big)\cdot \hat{B}_\alpha &= 0~,\quad\forall \alpha~\&~a~,
\end{align}
which identifies the internal states within a plaquette:
\begin{align}
    \adjincludegraphics[valign=c,scale=1]{ 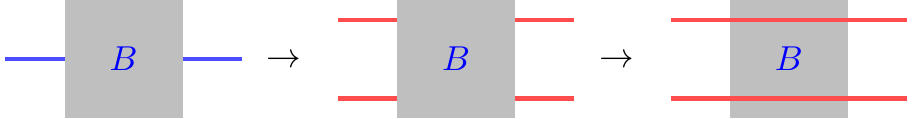}\,.
\end{align}
By further imposing charge neutral condition, we get solution for $\hat{B}_\alpha$ as
\begin{align}
    \hat{B}_\alpha= d_{e} \hat{b}_e + d_{o} \hat{b}_o\,,
\end{align}
where $\hat{b}_e = \vbra{\uparrow\uparrow}_u\vbra{\uparrow\uparrow}_v + \vbra{\downarrow\downarrow}_u\vbra{\downarrow\downarrow}_v$, and $\hat{b}_o = {\vbra{\uparrow\downarrow}_u\vbra{\uparrow\downarrow}_v}c_{u}c_{v} - \vbra{\downarrow\uparrow}_u\vbra{\downarrow\uparrow}_v$.
Here, $d_{e/o}$ are real parameters. 
By performing gauge transformation, we can always absorb $d$'s to site tensors and simply set $d_e=d_o=1$.
Namely, bond tensors are maximal entangled states, which share the same form as Eq.~(12) in the main text.

\section{Numerical calculation of many-body topological invariants}
\label{app:num}
In this appendix, we present the ansatz tensor for the honeycomb lattice with a bond dimension $4$ and perform explicit numerical calculations of the many-body topological invariant proposed by Shiozaki et al\cite{shiozaki2018many}. This analysis demonstrates the existence of a parameter region corresponding to the Quantum Spin Hall (QSH) phase.

By solving the symmetry and $\IGG$ restrictions (Eq.~\eqref{eq:teq_example2} and \eqref{eq:iga_example2}), we get ansatz tensors. The $u$-site variational tensor can be written as:
\begin{align}
    \hat{T}_{u}=\sum_{l} c^{u}_{l} t^{u}_{l}
\end{align} 
where $c^{u}_{l}$'s are real numbers, and $\hat{t}^{u}_{l}$'s are shown graphically below:
\begin{align*}
&\adjincludegraphics[valign=c]{ 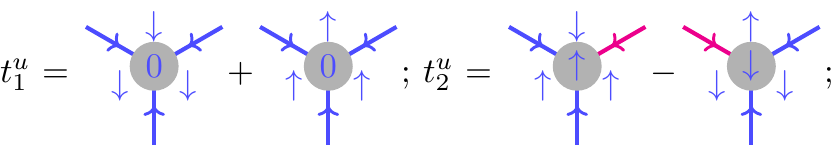}~\adjincludegraphics[valign=c]{ 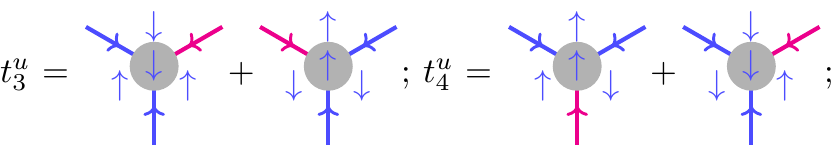}\\
    &\adjincludegraphics[valign=c]{ 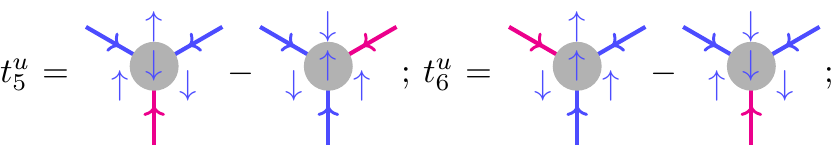}~\adjincludegraphics[valign=c]{ 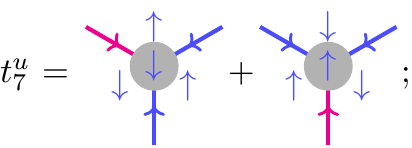}\\
\end{align*}
For $v$-site variational tensor, we have
\begin{align}
\hat{T}_{v}=\sum_{l}c^{v}_{l}t^{v}_{l}
\end{align}
with $c^{v}_l$'s real numbers and $\hat{t}_n$'s below
\begin{align*}
    &\adjincludegraphics[valign=c]{ 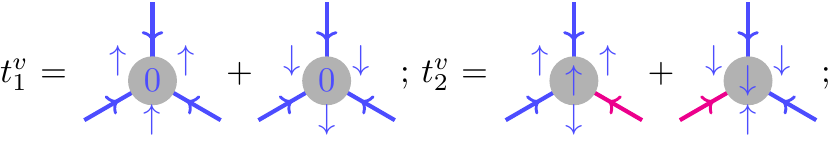}~\adjincludegraphics[valign=c]{ 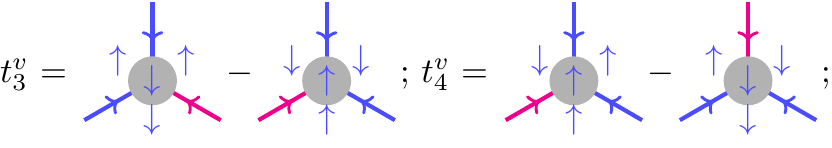}\\
    &\adjincludegraphics[valign=c]{ 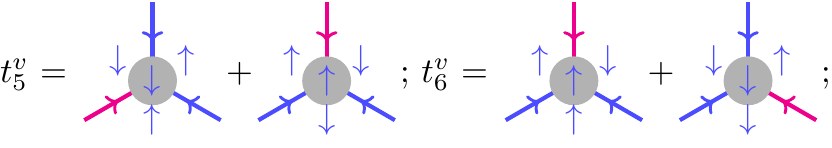}~\adjincludegraphics[valign=c]{ 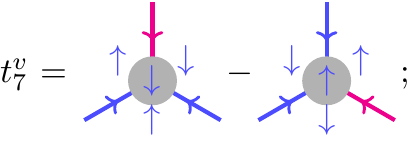}\\
\end{align*}

And bond tensor is chosen to be $\hat{B}=\vbra{\uparrow\uparrow}_u\vbra{\uparrow\uparrow}_v + \vbra{\downarrow\downarrow}_u\vbra{\downarrow\downarrow}_v-{\vbra{\uparrow\downarrow}_u\vbra{\uparrow\downarrow}_v}c_{u}c_{v} + \vbra{\downarrow\uparrow}_u\vbra{\downarrow\uparrow}_v$.
In contrast to the fixed-point wavefunction discussed in the main text, the variational wavefunction remains a superposition of loop configurations with decorated fermions. However, it now exhibits distinct coefficients and does not involve any additional physical Ising spins. Notably, the spin degrees of freedom become decoupled from the fermion degrees of freedom, existing solely as internal degrees of freedom.

To calculate the topological invariant, we set the $y$ direction of our 2+1D wavefunction to be periodic, and keep the $x$ direction open to get a cylindrical geometry. Then we divide the cylinder as below
\begin{align*}
    \adjincludegraphics[valign=c,scale=0.4]{fig/cylinder_geometry.pdf}
\end{align*}

The topological invariant of state $|\phi\rag$ is obtained from the following formula:
\begin{align}
    Z&=\Tr[\rho^{+}_{R_{1}\cup R_{3}} C^{R_{1}}_{T}(\rho^{-}_{R_{1}\cup R_{3}})^{\mathsf{T}_{1}}[C^{R_{1}}_{T}]^{\dagger}]\nonumber\\
    \rho^{\pm}_{R_{1}\cup R_{3}}&=\Tr_{\overline{R_{1}\cup R_{3}}}[\exp{\frac{\pm2\pi \mathrm{i}y\sum_{\mathrm{r}\in R_{2}}n(\mathrm{r})}{L_{y}}} |\phi\rag \lag \phi|]
\end{align}
Where $C^{R_{1}}_{T}(\rho^{-}_{R_{1}\cup R_{3}})^{\mathsf{T}_{1}}[C^{R_{1}}_{T}]^{\dagger}$ is the time-reversal partial transpose of $\rho^{-}_{R_{1}\cup R_{3}}$. In Fock space, the time-reversal partial transpose of $|\{n_{j}\}_{j\in R_{1}},\{n_{j}\}_{j\in R_{2}}\rag \lag \{\Bar{n}_{j}\}_{j\in R_{1}},\{\Bar{n}_{j}\}_{j\in R_{2}}|$ is defined as\cite{shiozaki2018many}
\begin{align}
    (-i)^{[\tau_{1}+\Bar{\tau}_{1}]} (-1)^{(\tau_{1}+\Bar{\tau}_{1})(\tau_{2}+\Bar{\tau_{2}})} U^{R_1}_{T}|\{\Bar{n}_{j}\}_{j\in R_{1}},\{n_{j}\}_{j\in R_{2}}\rag \lag \{n_{j}\}_{j\in R_{1}},\{\Bar{n}_{j}\}_{j\in R_{2}}|[U^{R_1}_{T}]^{\dagger}
\end{align}
with $[x]=0$ for even $x$ and $[x]=1$ for odd $x$ and $\tau_{1/2}=\sum_{j\in R_{1/2}} n_{j}$. $U^{R_{1}}_{T}$ is the unitary part of time-reversal action on region $R_{1}$.
The angle of $Z$ is $\pi$ for QSH phase and $0$ for trivial phase.

\begin{figure}[htpb]
    \centering
    \includegraphics[scale=1]{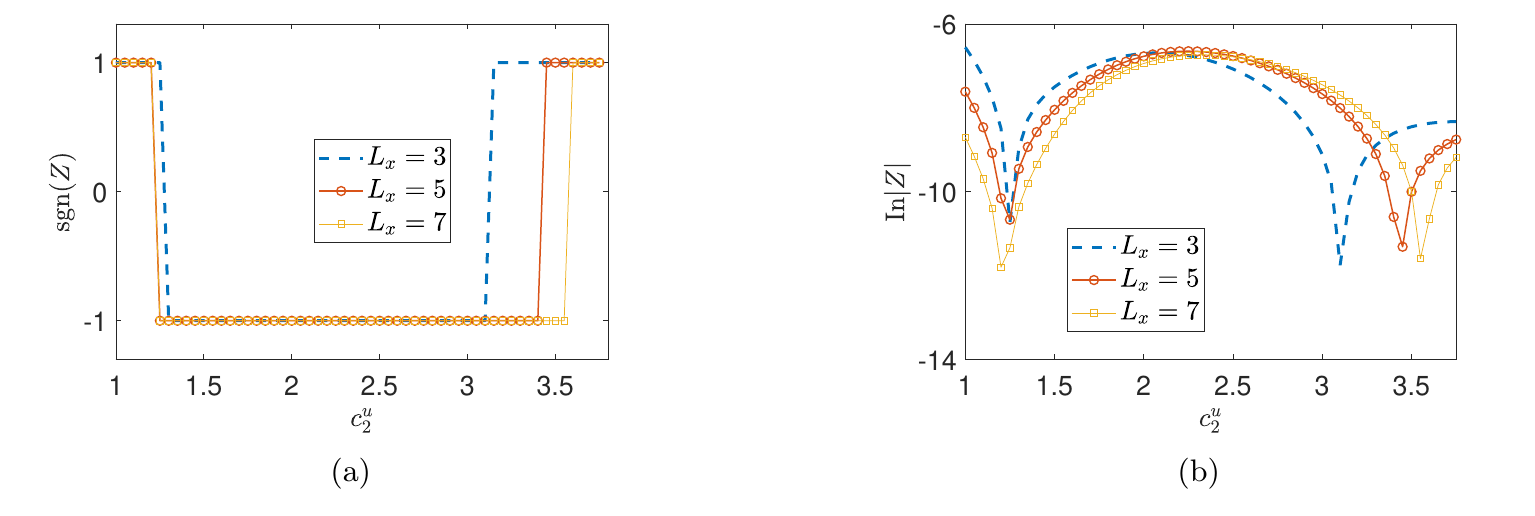}
    \caption{The sign (a) and amplitude (b) of $Z$ versus $c^{u}_{2}$ while other coefficients are set to be 1.}
    \label{fig:Z_cu2}
\end{figure}

We set the length of the $y$-direction, $L_y=2$ (two unit cells with four sites in the $y$-direction). Additionally, we set the length of three middle regions in the $x$-direction to be $L_x$. By tuning $c^{u}_{2}$ while keeping other coefficients to be $1$, for different $L_x$, we get the sign and amplitude of $Z$ as shown in Fig.~\ref{fig:Z_cu2}.

\begin{figure}[htpb]
    \centering
    \includegraphics[scale=1]{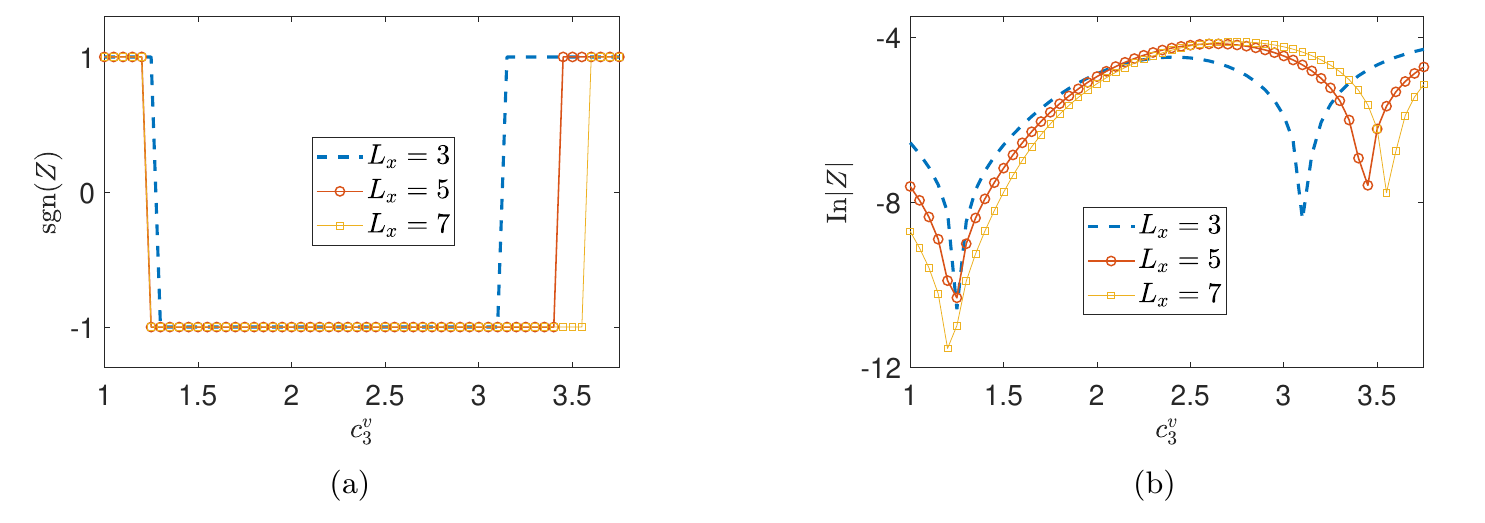}
    \caption{The sign (a) and amplitude (b) of $Z$ versus $c^{v}_{3}$ while other coefficients are set to be 1.}
    \label{fig:Z_cv3}
\end{figure}

\begin{figure}[htpb]
    \centering
    \includegraphics[scale=1]{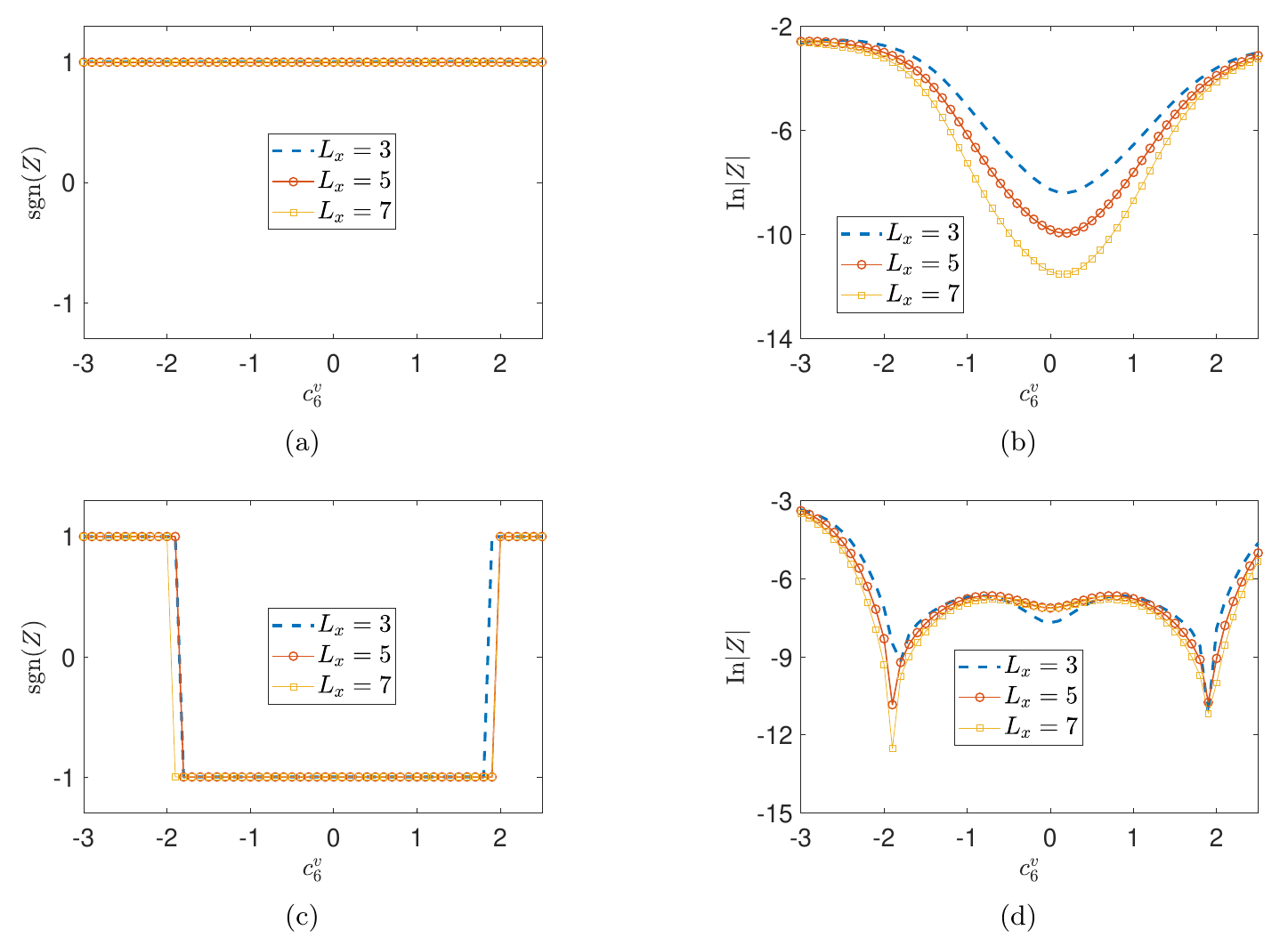}
    \caption{The sign (a)/(c) and amplitude (b)/(d) of $Z$ versus $c^{v}_{6}$ when $c^{u}_2=1$/$c^{u}_2=2$. Other coefficients are set to be $1$.}
    \label{fig:Z_cv6}
\end{figure}

We can also enter QSH phase by tuning other coefficients, like $c^{v}_{3}$, while keeping other coefficients to be $1$ as shown in Fig.~\ref{fig:Z_cv3}. Tuning other coefficients by the same means may not give us QSH phase region for the wavefunction are already out of QSH phase and those coefficients are irrelevant. But if we allow two coefficients to vary simultaneously, it is possible to make every coefficients relevant. To see this, we compare the sign of $Z$ versus $c^{v}_{6}$ and different $c^{u}_{2}$ while keeping other coefficients to be $1$. As shown in Fig.\ref{fig:Z_cv6} , we can't find QSH phase region when $c^{u}_{2}$ and other coefficients are $1$, but it is possible when $c^{u}_{2}=2$. Consequently, there is a wide parameter region in QSH phase that can be tuned for variational simulations.

The results show that there exists a stable parameter region of QSH phase when the system size is large enough. The emergence of the trivial phase region can be attributed to two factors. Firstly, it is influenced by the limitation of system size, as finite-size effects become significant beyond the fixed point. Secondly, our ansatz possesses the capability to represent long-range entangled GHZ (Greenberger-Horne-Zeilinger) states because emergent $\IGG$ cannot be ruled out solely based on local tensor equations.

\end{document}